\documentclass{aa}
\usepackage{graphicx}
\usepackage[varg]{txfonts}
\usepackage{hyperref}
\usepackage{ulem}
\usepackage{bm}
\usepackage{amsmath}
\usepackage{physics}
\usepackage{upgreek}
\usepackage{mathtools}
\usepackage{natbib}
\usepackage{float}
\usepackage{subfigure}
\bibpunct{(}{)}{;}{a}{}{,} % to follow the A&A style
\usepackage{CJK}
\hypersetup{
    colorlinks=true,
    linkcolor=blue,
    filecolor=magenta,  
    citecolor=blue,
    urlcolor=cyan,
    pdfpagemode=FullScreen,
    }
\def\setsymbol#1#2{\expandafter\def\csname #1\endcsname{#2}}
\def\getsymbol#1{\csname #1\endcsname}

\newbox\tablebox    \newdimen\tablewidth
\def\leaderfil{\leaders\hbox to 5pt{\hss.\hss}\hfil}
\def\tablenote#1 #2\par{\begingroup \parindent=0.8em
    \abovedisplayshortskip=0pt\belowdisplayshortskip=0pt
    \noindent
    $$\hss\vbox{\hsize\tablewidth \hangindent=\parindent \hangafter=1 \noindent
    \hbox to \parindent{$^#1$\hss}\strut#2\strut\par}\hss$$
    \endgroup}

\def\L2{\ifmmode L_2\else $L_2$\fi}

\def\DeltaT{\ifmmode \Delta T\else $\Delta T$\fi}
\def\deltat{\ifmmode \Delta t\else $\Delta t$\fi}
\def\fknee{\ifmmode f_{\rm knee}\else $f_{\rm knee}$\fi}
\def\Fmax{\ifmmode F_{\rm max}\else $F_{\rm max}$\fi}
\def\solar{\ifmmode{\rm M}_{\mathord\odot}\else${\rm M}_{\mathord\odot}$\fi}
\def\Msolar{\ifmmode{\rm M}_{\mathord\odot}\else${\rm M}_{\mathord\odot}$\fi}
\def\Lsolar{\ifmmode{\rm L}_{\mathord\odot}\else${\rm L}_{\mathord\odot}$\fi}
\def\inv{\ifmmode^{-1}\else$^{-1}$\fi}
\def\mo{\ifmmode^{-1}\else$^{-1}$\fi}
\def\sup#1{\ifmmode ^{\rm #1}\else $^{\rm #1}$\fi}
\def\expo#1{\ifmmode \times 10^{#1}\else $\times 10^{#1}$\fi}
\def\,{\thinspace}
\def\lsim{\mathrel{\raise .4ex\hbox{\rlap{$<$}\lower 1.2ex\hbox{$\sim$}}}}
\def\gsim{\mathrel{\raise .4ex\hbox{\rlap{$>$}\lower 1.2ex\hbox{$\sim$}}}}

\def\simprop{\mathrel{\raise .4ex\hbox{\rlap{$\propto$}\lower 1.2ex\hbox{$\sim$}}}}
\def\deg{\ifmmode^\circ\else$^\circ$\fi}
\def\pdeg{\ifmmode $\setbox0=\hbox{$^{\circ}$}\rlap{\hskip.11\wd0 .}$^{\circ}
          \else \setbox0=\hbox{$^{\circ}$}\rlap{\hskip.11\wd0 .}$^{\circ}$\fi}
\def\arcs{\ifmmode {^{\scriptstyle\prime\prime}}
          \else $^{\scriptstyle\prime\prime}$\fi}
\def\arcm{\ifmmode {^{\scriptstyle\prime}}
          \else $^{\scriptstyle\prime}$\fi}
\newdimen\sa  \newdimen\sb
\def\parcs{\sa=.07em \sb=.03em
     \ifmmode \hbox{\rlap{.}}^{\scriptstyle\prime\kern -\sb\prime}\hbox{\kern -\sa}
     \else \rlap{.}$^{\scriptstyle\prime\kern -\sb\prime}$\kern -\sa\fi}
\def\parcm{\sa=.08em \sb=.03em
     \ifmmode \hbox{\rlap{.}\kern\sa}^{\scriptstyle\prime}\hbox{\kern-\sb}
     \else \rlap{.}\kern\sa$^{\scriptstyle\prime}$\kern-\sb\fi}
\def\ra[#1 #2 #3.#4]{#1\sup{h}#2\sup{m}#3\sup{s}\llap.#4}
\def\dec[#1 #2 #3.#4]{#1\deg#2\arcm#3\arcs\llap.#4}
\def\deco[#1 #2 #3]{#1\deg#2\arcm#3\arcs}
\def\rra[#1 #2]{#1\sup{h}#2\sup{m}}

\def\dots{\relax\ifmmode \ldots\else $\ldots$\fi}
\def\WHzsr{\ifmmode $W\,Hz\mo\,sr\mo$\else W\,Hz\mo\,sr\mo\fi}
\def\mHz{\ifmmode $\,mHz$\else \,mHz\fi}
\def\GHz{\ifmmode $\,GHz$\else \,GHz\fi}
\def\mKs{\ifmmode $\,mK\,s$^{1/2}\else \,mK\,s$^{1/2}$\fi}
\def\muKs{\ifmmode \,\mu$K\,s$^{1/2}\else \,$\mu$K\,s$^{1/2}$\fi}
\def\muKRJs{\ifmmode \,\mu$K$_{\rm RJ}$\,s$^{1/2}\else \,$\mu$K$_{\rm RJ}$\,s$^{1/2}$\fi}
\def\muKHz{\ifmmode \,\mu$K\,Hz$^{-1/2}\else \,$\mu$K\,Hz$^{-1/2}$\fi}
\def\MJysr{\ifmmode \,$MJy\,sr\mo$\else \,MJy\,sr\mo\fi}
\def\MJysrmK{\ifmmode \,$MJy\,sr\mo$\,mK$_{\rm CMB}\mo\else \,MJy\,sr\mo\,mK$_{\rm CMB}\mo$\fi}
\def\microns{\ifmmode \,\mu$m$\else \,$\mu$m\fi}

\def\muK{\ifmmode \,\mu$K$\else \,$\mu$\hbox{K}\fi}
\def\microK{\ifmmode \,\mu$K$\else \,$\mu$\hbox{K}\fi}
\def\muW{\ifmmode \,\mu$W$\else \,$\mu$\hbox{W}\fi}
\def\kms{\ifmmode $\,km\,s$^{-1}\else \,km\,s$^{-1}$\fi}
\def\kmsMpc{\ifmmode $\,\kms\,Mpc\mo$\else \,\kms\,Mpc\mo\fi}
\providecommand{\sorthelp}[1]{}

\begin{document}

\title{Implications of scattering for CMB foreground emission modelling}

\begin{CJK*}{UTF8}{gbsn}
\author{Jia-Rui Li (李嘉睿)
%    \thanks{\email{jr981025@mail.ustc.edu.cn}}
    \inst{\ref{ustc1},\ref{ustc2},\ref{ustc3}}
\and Jacques Delabrouille
    \thanks{\email{delabrouille@apc.in2p3.fr}} \inst{\ref{cpb},\ref{lbl}}
\and Yi-Fu Cai (蔡一夫)
    \thanks{\email{yifucai@ustc.edu.cn}} \inst{\ref{ustc1},\ref{ustc2},\ref{ustc3}}
\and Dongdong Zhang (张东东)
    \inst{\ref{ustc1},\ref{ustc2},\ref{ustc3},\ref{ipmu}}
    }

\institute{
Department of Astronomy, School of Physical Sciences, University of Science and Technology of China, Hefei 230026, China \label{ustc1}
\and School of Astronomy and Space Science, University of Science and Technology of China, Hefei, Anhui 230026, China \label{ustc2}
\and CAS Key Laboratory for Researches in Galaxies and Cosmology, University of Science and Technology of China, Hefei 230026, China \label{ustc3}
\and CNRS-UCB International Research Laboratory, Centre Pierre Bin\'etruy, IRL 2007, CPB-IN2P3, Berkeley, CA 94720, USA \label{cpb} 
\and Lawrence Berkeley National Laboratory, 1 Cyclotron Road, Berkeley, CA 94720, USA \label{lbl}
\and Kavli IPMU (WPI), UTIAS, The University of Tokyo, Kashiwa, Chiba 277-8583, Japan \label{ipmu}}

\date{Received ; accepted }

% context, aims, methods, results, conclusions
\abstract
{The extreme precision and accuracy of forthcoming observations of CMB temperature and polarization anisotropies, aiming to detect the tiny signatures of primordial gravitational waves or of light relic particles beyond the standard three light neutrinos, requires commensurate precision in the modelling of foreground Galactic emission that contaminates CMB observations.}
{We evaluate the impact of second-order effects in Galactic foreground emission due to Thomson scattering off interstellar free electrons and to Rayleigh scattering off interstellar dust particles. }
{We use existing sky survey data and models of the distribution of free electrons and dust within the Milky Way to estimate the amplitude and power spectra of the emission originating from radiation scattered either by free electrons or by dust grains at CMB frequencies. }
{Both processes generate corrections to the total emission that are small compared to direct emission, and are small enough not to pose problems for current-generation observations. }
{However, B-modes generated by Thomson scattering of incoming radiation by interstellar free electrons at CMB frequencies are within an order of magnitude of the sensitivity of the most advanced forthcoming CMB telescopes, and might require more precise evaluation in the future. }


\titlerunning{}
\authorrunning{Jia-Rui Li, et al.}
\maketitle
\end{CJK*}

\section{Introduction}
Since the discovery of the Cosmic Microwave Background (CMB) by Penzias and Wilson \citep{1965ApJ...142..419P}, the CMB has emerged as a powerful cosmological probe, enabling deeper understanding of the history and evolution of our universe, as well as rigorous testing of cosmological models. 

The WMAP \citep{bennett2012} and \textit{Planck} \citep{planck2016-l01} space missions have achieved remarkable precision in measuring the full-sky temperature anisotropies of the CMB, while also providing invaluable survey data concerning CMB polarization modes. 
Still, much information can yet be extracted from more sensitive and accurate observations of the CMB. This motivates robust ongoing ground-based, balloon-borne, and space-borne observational programs. 
Observations with experiments such as the POLARBEAR \citep{2010SPIE.7741E..1EA}, SPT \citep{2011PASP..123..568C}, BICEP-Keck experiment \citep{BicepDetection}, ACTPol \citep{2014JCAP...10..007N}, and CLASS \citep{2020JLTP..199..289D}, among others, have gradually improved the accuracy and sensitivity of polarization measurements over patches of the southern sky, paving the way for even more ambitious ground-based observations with AliCPT \citep{2020SPIE11453E..2AS}, the Simons Observatory \citep{2019JCAP...02..056A} and with CMB-S4 \citep{2019arXiv190704473A}, as well as with next-generation space missions such as LiteBIRD \citep{2023PTEP.2023d2F01L}, PICO \citep{2019arXiv190210541H}, CORE \citep{2018JCAP...04..014D}, PIXIE \citep{2011JCAP...07..025K} and PRISM \citep{2014JCAP...02..006A}.

One of the key objectives of these experiments is the detection, in CMB polarization patterns, of the tiny signature of primordial gravitational waves produced during a phase of cosmic inflation in the extremely early universe. 
These so-called primordial B-modes of CMB polarization are expected to be orders of magnitude fainter than foreground polarized emission from the Galactic interstellar medium (ISM), which has motivated numerous studies of the detectability of CMB B-modes in the presence of foreground contamination \citep{2005MNRAS.360..935T, 2009A&A...503..691B, 2009AIPC.1141..222D, 2016PhRvD..94h3526S, 2018JCAP...04..023R, 2019arXiv190508888T, 2023JCAP...06..034A, 2024arXiv240201233Z}. 
However, the conclusion of those studies typically depends on the reliability of the assumptions being made on the properties of the foreground emission.

It is generally assumed that two main components in the Galactic ISM contribute to the foreground polarization emission for CMB observations. 
Synchrotron emission, generated by relativistic electrons spiralling in the Galactic magnetic field, is strongly polarized, and dominates at low frequencies, below $\nu \simeq 80$~GHz, while thermal dust emission from elongated dust grains, which dominates above $\nu \simeq 80$~GHz, is polarized by reason of grain alignment perpendicular to the Galactic magnetic field.
Free-free emission from warm ionized medium, dust emission from spinning dust grains, and line emissions from atoms and molecules such as carbon monoxide, which significantly contribute to foreground emission in intensity, are usually assumed to be unpolarized, or very little polarized, and thus to negligibly impact the measurement of CMB polarization in low-foreground emission regions of the sky. 

With ever more precise CMB observations, it becomes crucial to thoroughly examine the possible interactions of these components and their impact on the detailed properties of polarized foreground emission. 
In general, free electrons and Galactic dust grains can absorb or scatter incoming radiation, potentially generating distortions in the model electromagnetic spectra by second order effects. 

For example, while primordial thermal free-free emission is unpolarized due to the isotropy of the random coulombian collisions between particles, \citet{2014PTEP.2014fB109I} pointed out that Thomson scattering by free electrons which generate the free-free emission will produce polarized photons. 
In another instance, fee electrons responsible for free-free emission can also absorb synchrotron radiation at low frequencies, around 408\,MHz, which may produce bias in the modelling of synchrotron radiation when extrapolated to higher frequency \citep{2008A&A...477..573S,delabrouille2012}. 
And also, dust particles absorb CMB photons, generating a ``CMB shadow'' that generates an effective (negative) component of polarized emission by interstellar dust \citep{2020ApJ...895L..21N}.

In this paper, we investigate two more of these second-order effects, and estimate their magnitudes and impacts for modelling foreground emission. 
Of specific interest are the scattering of dust and synchrotron emission by a distribution of thermal free electrons, as well as the scattering of synchrotron and free-free emission by dust particles. 
Indeed, these second-order effects would generate apparent emission processes with frequency scaling different from that of the intrinsic emission of those scatterers.

The rest of this paper is organised as follows: 
In Section \ref{Section_Thomson_scattering_off_free_electrons} we derive mathematical formulae and approximation for the scattering of Galactic emission by free electrons, and derive corresponding numerical estimates of the scattering of incoming radiation for a recent model of free-electron distribution. 
Section \ref{Section_Rayleigh_scattering_by_ISM_dust_grains} checks the amplitude of Rayleigh scattering caused by dust grains, and we conclude in Section \ref{Section_Conclusion}.

\section{Thomson scattering off free electrons}
\label{Section_Thomson_scattering_off_free_electrons}

Scattering of anisotropic radiation by free electrons generates polarized emission originating from the scatterers. 
This is the exact process that generates most of CMB polarization at the surface of last scattering. 
In our own Galaxy, free electrons are potential scatterers of incoming emission originating from the Milky Way. 
This may generate a component of polarized emission associated with the warm ionized medium in the galactic ISM. 

\subsection{Scattering model}
The intensity and polarization properties of an electromagnetic wave with frequency $\nu$ propagating along direction $\bm{n}$ can be totally quantified with a rank-2 tensor, the ``polarization tensor'', denoted by $\bm{J}_\nu$ 
%in units of $\mathrm{W\,m^{-2}\,Hz^{-1}\,sr^{-1}}$
\citep{Landau02}: 

\begin{equation}
\begin{aligned}
\bm{J}_\nu 
&= \frac{2}{\mu_0 c}\langle\bm{E}_\nu\otimes{\bm{E}_\nu}^*\rangle
= \sum_{i,j=1}^2 J_{\nu,ij}\,\,\bm{e}_i\otimes\bm{e}_j,
\label{Equation_Emission Polarization Tensor}
\end{aligned}
\end{equation}
where the complex vector $\bm{E}_\nu$ is the electric component of the electromagnetic wave and $\bm{E}_\nu^*$ is its complex conjugate. The triangular brackets indicate averaging over time.\footnote{
    The factor in Eq. (\ref{Equation_Emission Polarization Tensor}), $2/\mu_0 c$, can be determined as follows: according to Poynting's theorem, 
    the intensity of radiation, $\dd I_\nu$, into the infinitesimal solid angle element $\dd\Omega$ is
    \begin{equation}
    \dd I_\nu = 2\left\langle\left|\bm{S}_\nu\right|\right\rangle r^2 \dd\Omega 
    = \dfrac{2}{\mu_0}\left\langle\left|\bm{E}_\nu\times\bm{B}_\nu\right|\right\rangle r^2 \dd\Omega
    = \dfrac{2}{\mu_0 c}\left\langle\left|\bm{E}_\nu\right|^2\right\rangle r^2 \dd\Omega,
    \end{equation}
    where $\bm{S}_\nu$, $\bm{B}_\nu$, $\bm{E}_\nu$ represent the Poynting vector (and also the energy flux density of the electromagnetic wave), electric and magnetic vectors with frequency $\nu$, at a distance $r$ from the emission source, respectively. 
    The factor $2$ arises here due to the one-side Fourier transform in the range with positive frequency. 
    Thus the trace of the polarization tensor, that is, intensity per unit area should be
    \begin{equation}
    \frac{\dd I_\nu}{r^2\dd\Omega} = \frac{2}{\mu_0 c}\langle\bm{E}_\nu^*\cdot\bm{E}_\nu\rangle.
    \end{equation}
}
For brevity, the subscript $\nu$ denoting frequency will be omitted in the following. 
By selecting orthogonal unit vectors $\bm{e}_1$ and $\bm{e}_2$ perpendicular to the propagation direction $\bm{n}$ as the bases, the polarization tensor %is composed of components 
can be written as
\begin{equation}
J_{ij} = \frac{1}{2}
\begin{pmatrix}
I+Q & U-\mathrm{i}V\\
U+\mathrm{i}V & I-Q
\end{pmatrix},\qquad i,j = 1,2
\end{equation}
where $I$, $Q$, $U$ and $V$ are Stokes parameters corresponding to $\bm{n}$. 

Most free electrons in the Galactic ISM are in the form of warm ionized medium (WIM), and emit free-free emission that is essentially unpolarized.
Thomson scattering of incoming radiation by those free electrons, however, would generate additional apparent radiation originating from the WIM. 
The scattering of emission originating from our own Galaxy is of specific interest: incoming anisotropic emission (strong parallel to the Galactic plane, in particular from the Galactic centre, and faint in the orthogonal direction) will generate polarized scattering light with frequency scaling similar to that of the incoming total emission, scattered towards the observer from the direction of ionized bubbles. 
In this section, we estimate the magnitude and polarization level of this scattered emission. 

%Thermal dust emission is usually modelled with one or more modified blackbody emission laws, in the form 
%\begin{equation}
%\bm{s}_{\mathrm{dust}} \propto \bm{a}_{\mathrm{dust}} \nu^{\beta_d} B_\nu(T_d).
%\end{equation}

Free electrons, which are responsible for free-free emission in the Milky Way, exhibit a characteristic temperature of approximately $8000$~K \citep{2001RvMP...73.1031F}. 
Their thermal velocities being tiny compared to the speed of light, the scattering of light off these electrons falls within the regime of Thomson scattering, and 
the electric vector of the scattered light, $\bm{E}'$, is related to both the incident light's electric vector $\bm{E}$ and the direction of scattering $\bm{n}'$ (with the prime $'$ denoting scattered quantities hereafter) according to 
\begin{equation} 
\bm{E}' = \frac{\mu_0 q^2}{4\uppi m_e r'} \bm{n}' \times \left(\bm{n}' \times \bm{E}\right)
= \frac{\mu_0 q^2}{4\uppi m_e r'} \Big[\big(\bm{n}'\cdot\bm{E}\big)\bm{n}' - \bm{E}\Big],
\end{equation}
where $r'$ is the distance between the electron and the observer, $m_e$ is the electron mass, and $q = 1.602\times10^{-19}\,\mathrm{C}$ is the elementary charge. 

\begin{figure}
\centering
\includegraphics[width=0.49\textwidth]{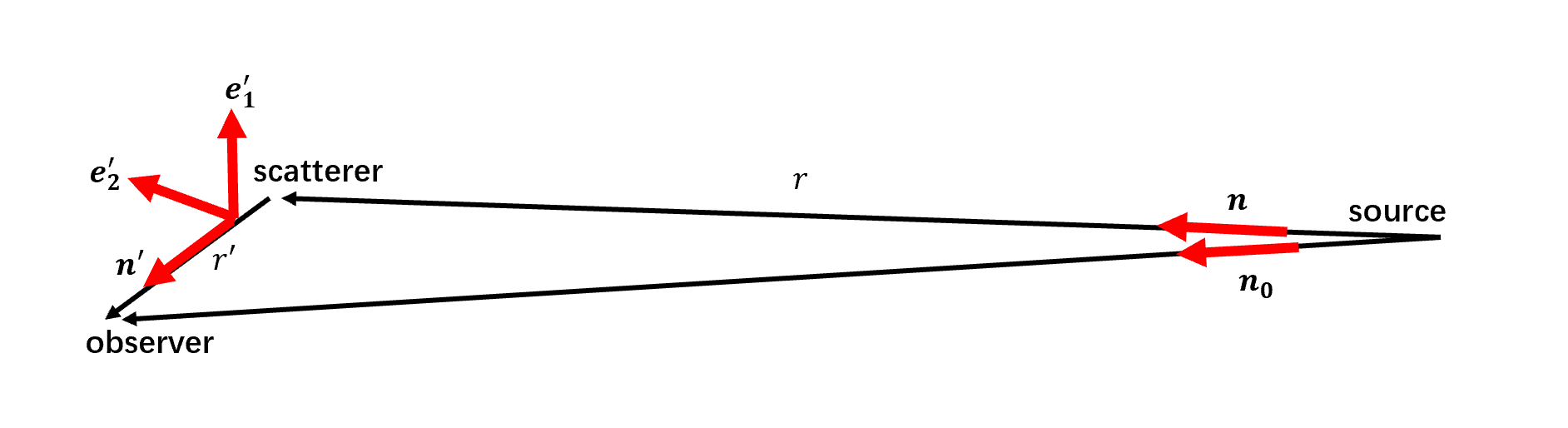}
\caption{Sketch depicting the relative positions of the emission source, scatterer and observer. The unit direction vectors from the source to the scatterer, from the scatterer to the observer, and from the source to the observer are represented by $\bm{n}$, $\bm{n}'$ and $\bm{n}_0$ respectively. 
The distance between the source and the scatterer is denoted as $r$, while $r'$ corresponds to the distance between the scatterer and the observer. 
Additionally, $\bm{e}'_1$ and $\bm{e}'_2$ serve as the reference base vectors for polarization. 
The set of vectors $(\bm{e}'_1,\,\bm{e}'_2,\,\bm{n}')$ form a right-hand coordinate system. }
\label{Figure_sketch_map}
\end{figure}

The polarization characteristics of the scattered light are determined by the projection of the polarization tensor onto the two base vectors $\bm{e}'_1$ and $\bm{e}'_2$ perpendicular to the scattering direction: 
\begin{equation}
\bm{E}' \cdot \bm{e}'_k 
= -\frac{\mu_0 q^2}{4\uppi m_e r'} \big(\bm{E}\cdot\bm{e}'_k\big), \qquad k = 1, 2.
\end{equation}

For CMB observations, we are mostly interested in foreground emission at high Galactic latitude, which mainly originates from the vicinity of Earth in our Galaxy. Indeed, the number density of free electrons decays rapidly as they move away from the Galactic plane \citep{1989ApJ...339L..29R}. 
Thus it is mostly free electrons near Earth that produce most of the scattering in mid-high latitude. 
For those free electrons, the integrated Galactic emission as a function of direction is similar to that observed on Earth, and as a first approximation we can represent the incoming radiation with a single two-dimensional background map, identical for all scatterers, and identical to what is directly observed with telescopes on Earth or in orbit. 
This is, however, only an approximation, which may not hold in local regions of strong inhomogeneity of the local ISM, and thus of the local incoming radiation. We just neglect this subtlety in the present work.

A free electron in direction $-\bm{n}'$, with distance $r'$ to Earth, receives incident light from all directions, and then scatters them towards all directions, and particularly in direction $\bm{n}'$, i.e., towards us. 
Thus the component $J'_{0,ab}$ ($a, b = 1, 2$) of the scattered light by this single electron,\footnote{The subscript 0 represents the quantity corresponding to a single electron. }
representing the result of the polarization tensor of the scattered light operating on (i.e., being projected onto) the base vectors $\bm{e}'_a$ and $\bm{e}'_b$, should be 
\begin{equation}
\begin{aligned}
J'_{0, ab} & = \frac{2}{\mu_0 c}\Big\langle\big(\bm{E}'\cdot\bm{e}'_a\big)\big({\bm{E}'}^*\cdot\bm{e}'_b\big)\Big\rangle\\
&= \frac{3}{8\uppi} \frac{\sigma_T}{{r'}^2}\cdot \frac{2}{\mu_0 c} \int \Big\langle\big(\bm{E}\cdot\bm{e}'_a\big) \big(\bm{E}^*\cdot\bm{e}'_b\big)\Big\rangle \,\dd\Omega,
\end{aligned}
\end{equation}
where $\sigma_T = \dfrac{8\uppi}{3}\left(\dfrac{q^2}{4\mathrm{\pi}\varepsilon_0 m c^2}\right)^2 = 6.653\times10^{-29}\,\mathrm{m^2}$ is the Thomson scattering cross-section. 
The integration over solid angle $\dd\Omega$ is due to the fact that the electron scatters a fraction of the incident radiation $\bm{E}$ from all incoming directions towards any specific direction (here, towards us). 

Along a certain line of sight defined by the direction vector $\bm{n}'$, we integrate over the infinitesimal volume element $\displaystyle {r'}^2 \dd r' \dd\Omega'$ containing free electrons to obtain the scattered signal power per unit area, per unit solid angle, and per unit frequency interval measured on Earth, 
\begin{equation}
\begin{aligned}
J'_{ab}(\bm{n}') & = \frac{1}{\dd\Omega'}\int J'_{0, ab}\, N_\mathrm{e}{r'}^2 \dd r'\,\dd\Omega'\\
&\simeq \frac{3}{8\uppi} \sigma_T \mathrm{DM}(\bm{n}') \sum_{i,j=1}^2\int J_{ij}(\bm{n})\big(\bm{e}_i\cdot\bm{e}'_a\big) \big(\bm{e}_j\cdot\bm{e}'_b\big) \,\dd\Omega,
\end{aligned}
\label{Equation_approximate observed polarization tensor}
\end{equation}
where $\mathrm{DM}(\bm{n}') \equiv \displaystyle\int N_\mathrm{e}\dd r'$ is the dispersion measure of free electrons along direction $\bm{n}'$, and $N_\mathrm{e}$ is the number density of free electrons. 
Due to the exponential decay of the number density of free electrons with distance from the Galactic plane, we make the approximation that the vast majority of scattering electrons at mid to high Galactic latitudes are concentrated close to Earth. 
Therefore, in Eq. (\ref{Equation_approximate observed polarization tensor}), one can separate the integration over solid angle and the integration over distance along the line of sight. 

\subsection{Numerical simulations}

\begin{figure}
\centering
\includegraphics[width=0.49\textwidth]{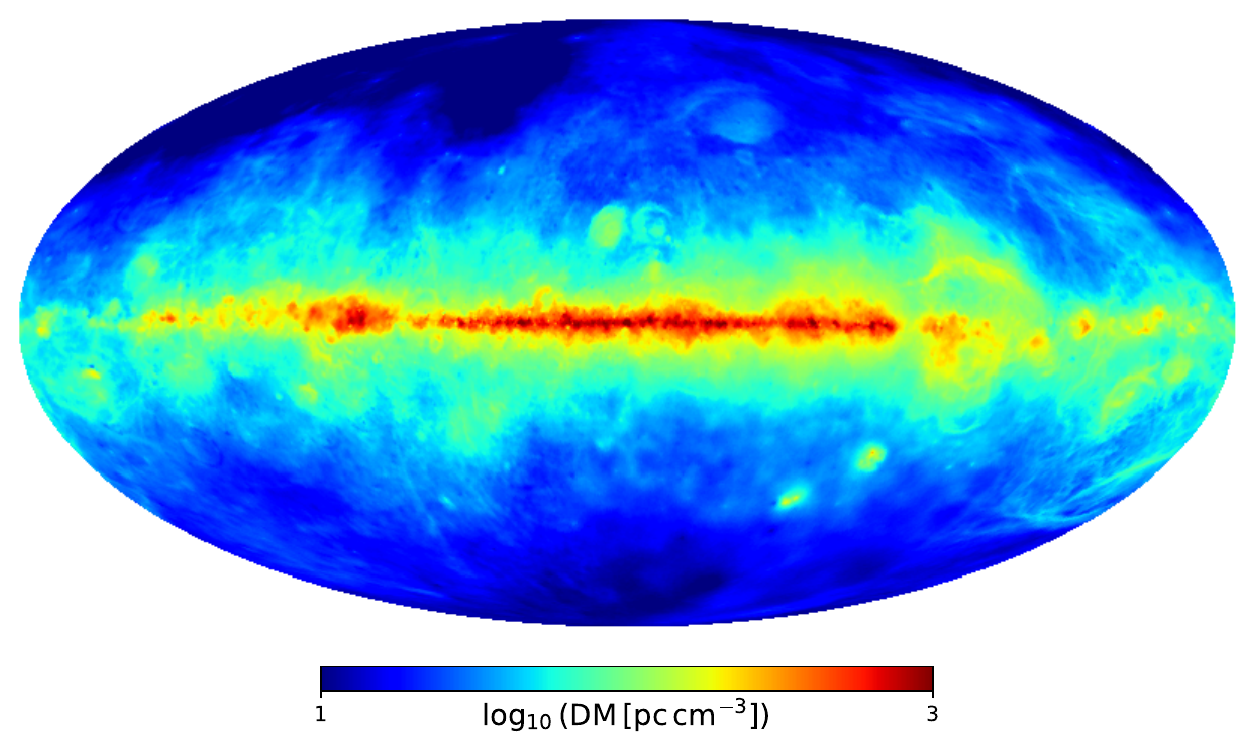}
\caption{Dispersion measure of Galactic electrons in logarithmic colour mapping, from \citet{2023arXiv230412350H}}
\label{Figure:DM}
\end{figure}

We now turn to numerical estimations of scattered foreground emission from the ionized interstellar medium as observable at millimetre wavelengths for CMB experiments, with an emphasis on emission from high Galactic latitude. 
Although a complete and accurate calculation of the scattered light requires a complete three-dimensional model of the distribution of the free electrons and of radiation at all positions, a simplified model can give an estimation of the amplitude and polarization of the scattered emission. 

According to Eq. (\ref{Equation_approximate observed polarization tensor}), this requires both a template for incoming radiation, and the dispersion measure of free electrons. 
The \textit{Planck} space mission \citep{planck2016-XLVIII} published comprehensive full-sky emission maps that can be used as templates for the incoming radiation in a set of frequency bands of interest. 
The dispersion measure of free electrons over the full-sky, however, is not very well known. 
It is measured through the dispersion of the arrival time of electromagnetic pulses emitted by pulsars as they traverse through electron-rich regions \citep{1993ApJ...411..674T}. 
\citet{2023arXiv230412350H} obtained a full-sky map of Galactic electron dispersion measure using pulsar dispersion measures (DMs) from the Australia Telescope National Facility catalogue \citep{2005AJ....129.1993M}, 
extragalactic Faraday rotation measures (RMs) from \citet{2023ApJS..267...28V}, 
emission measure (EM) map of Galactic electrons from \citet{planck2014-a12}, 
H-$\alpha$ emission map from \citet{2003ApJS..146..407F}, 
and three-dimensional distribution model of Galactic electrons by \citet[][hereafter YMW16]{2017ApJ...835...29Y}. 

The YMW16 three-dimensional model for thermal electron distribution only accounts for the largest-scale distribution and a few local structures, in particular the Gum Nebula and Local Cavity.
The DM map from \citet{2023arXiv230412350H}\footnote{DM\underline{ }mean\underline{ }std.fits downloaded from  \url{https://zenodo.org/records/10736552}}, shown in Fig. \ref{Figure:DM}, however, comprises smaller-scale fluctuations of electron number density. 
The angular power spectrum corresponding to the projection of these perturbations on the two-dimensional celestial sphere is compatible with the predictions based on the log-normal distribution for density perturbations following a Kolmogorov power spectrum in three-dimensional space \citep{2010ApJ...710..853C, 2013MNRAS.435.1610P}: 
\begin{equation}
P(k) \propto k^{-11/3},\qquad 10^6\,\mathrm{m} \lesssim \frac{2\uppi}{k} \lesssim 10^{18}\,\mathrm{m}.
\end{equation}

\begin{figure*}
\centering
\includegraphics[width=0.33\textwidth]{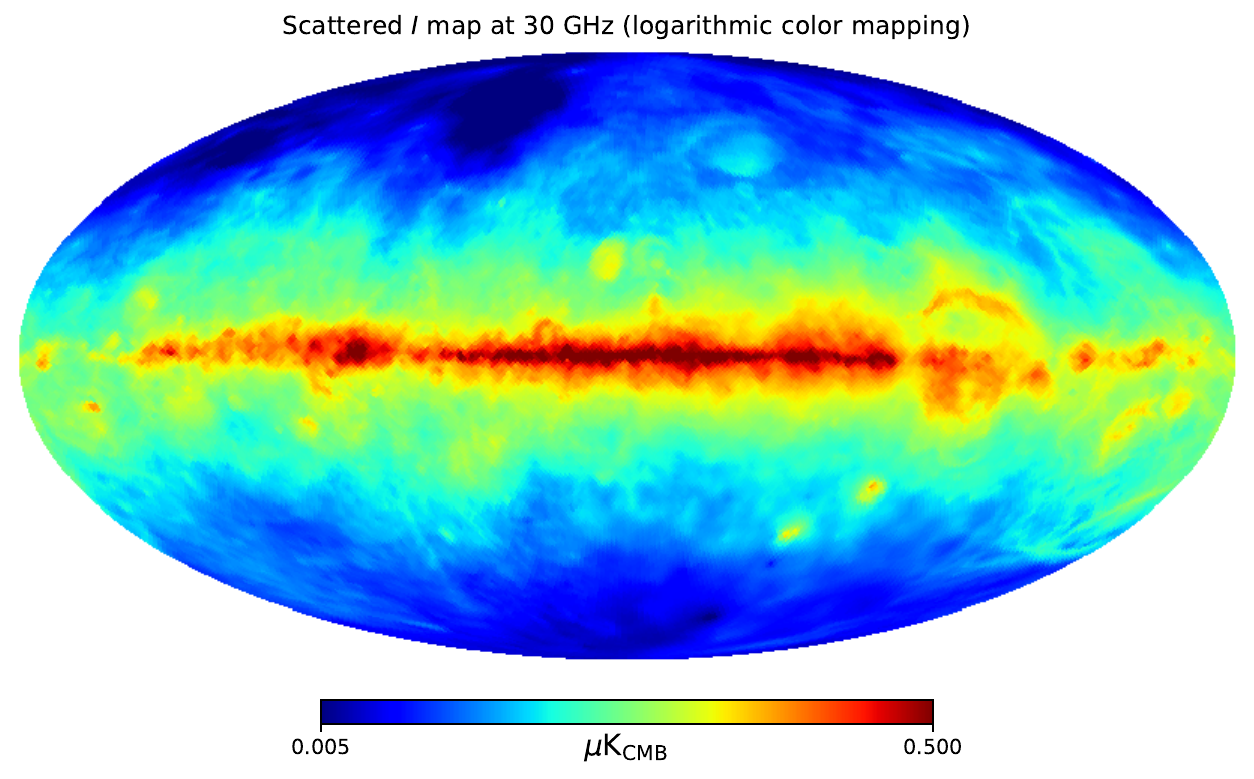}
\includegraphics[width=0.33\textwidth]{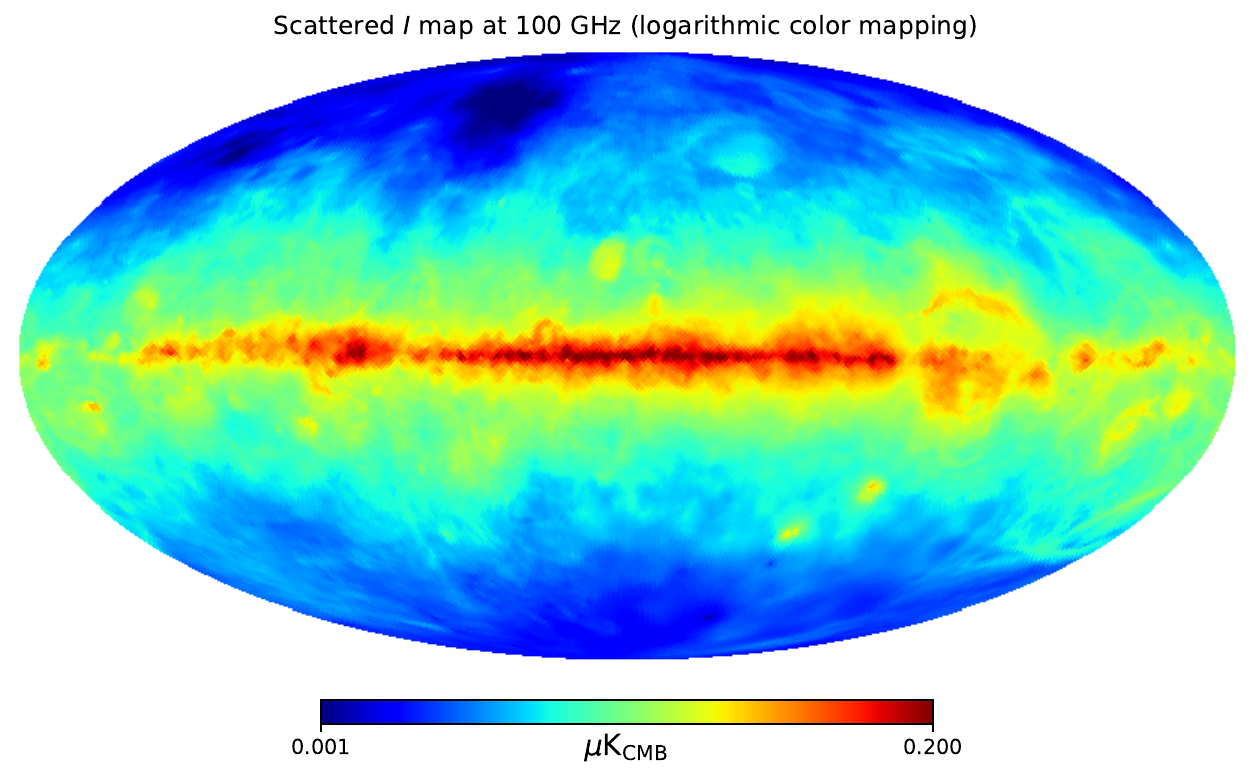}
\includegraphics[width=0.33\textwidth]{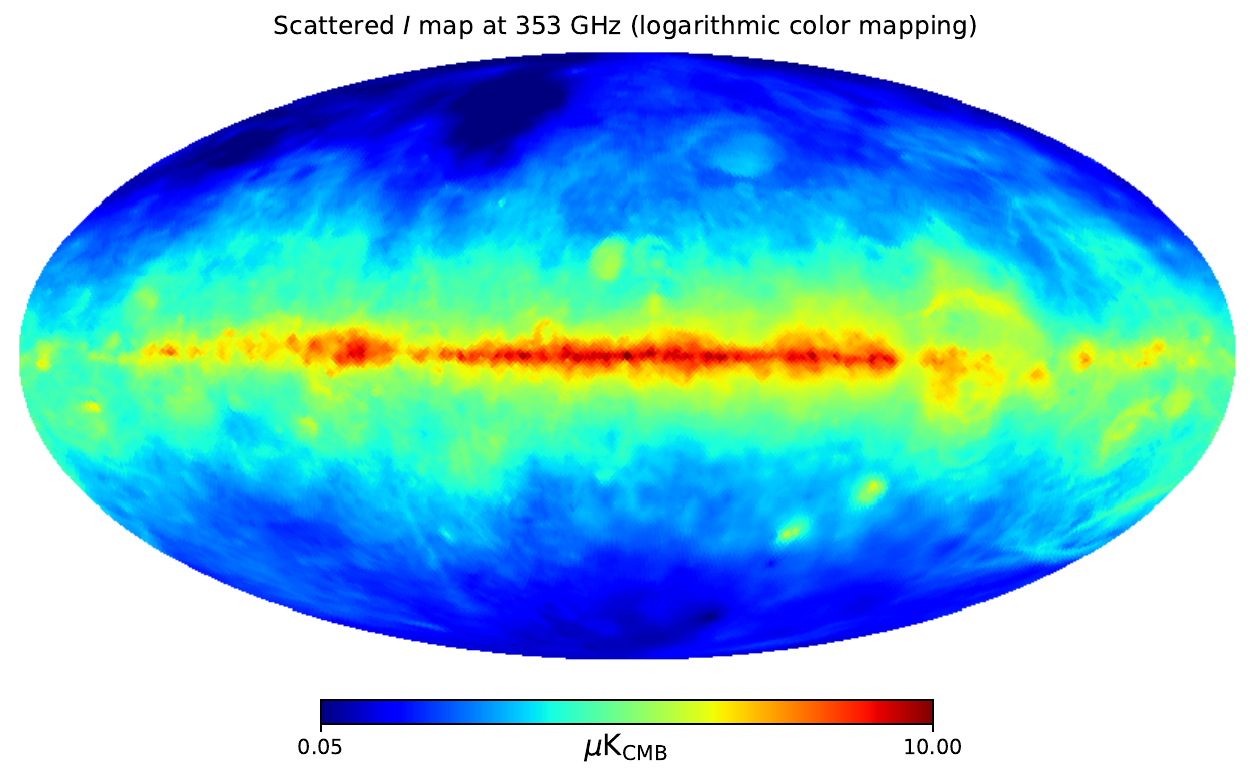}\\
\includegraphics[width=0.33\textwidth]{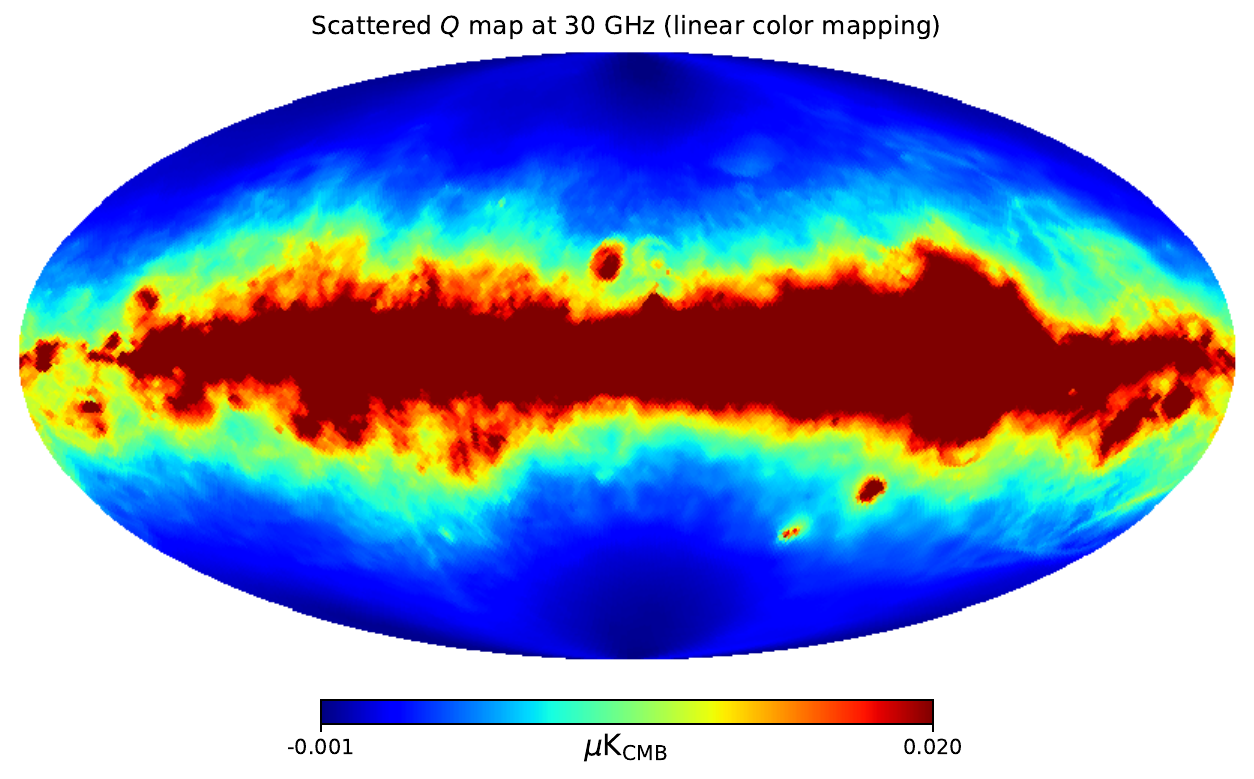}
\includegraphics[width=0.33\textwidth]{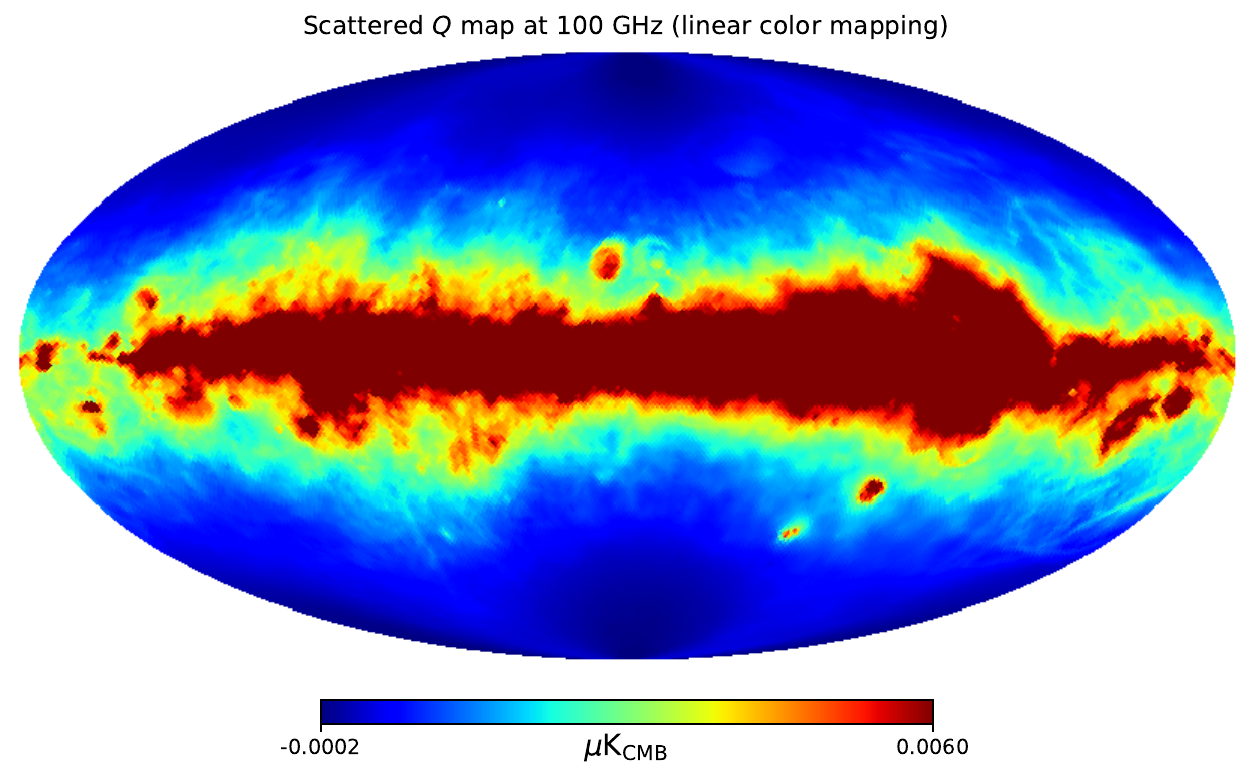}
\includegraphics[width=0.33\textwidth]{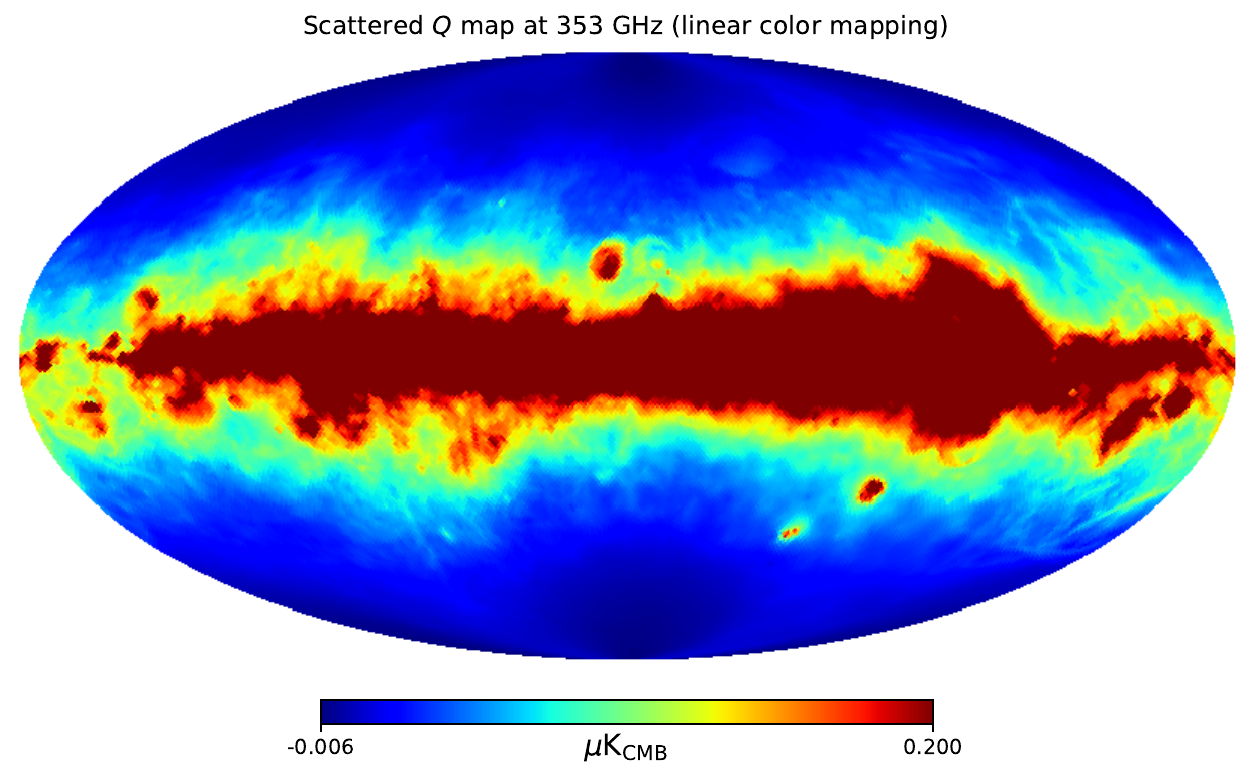}\\
\includegraphics[width=0.33\textwidth]{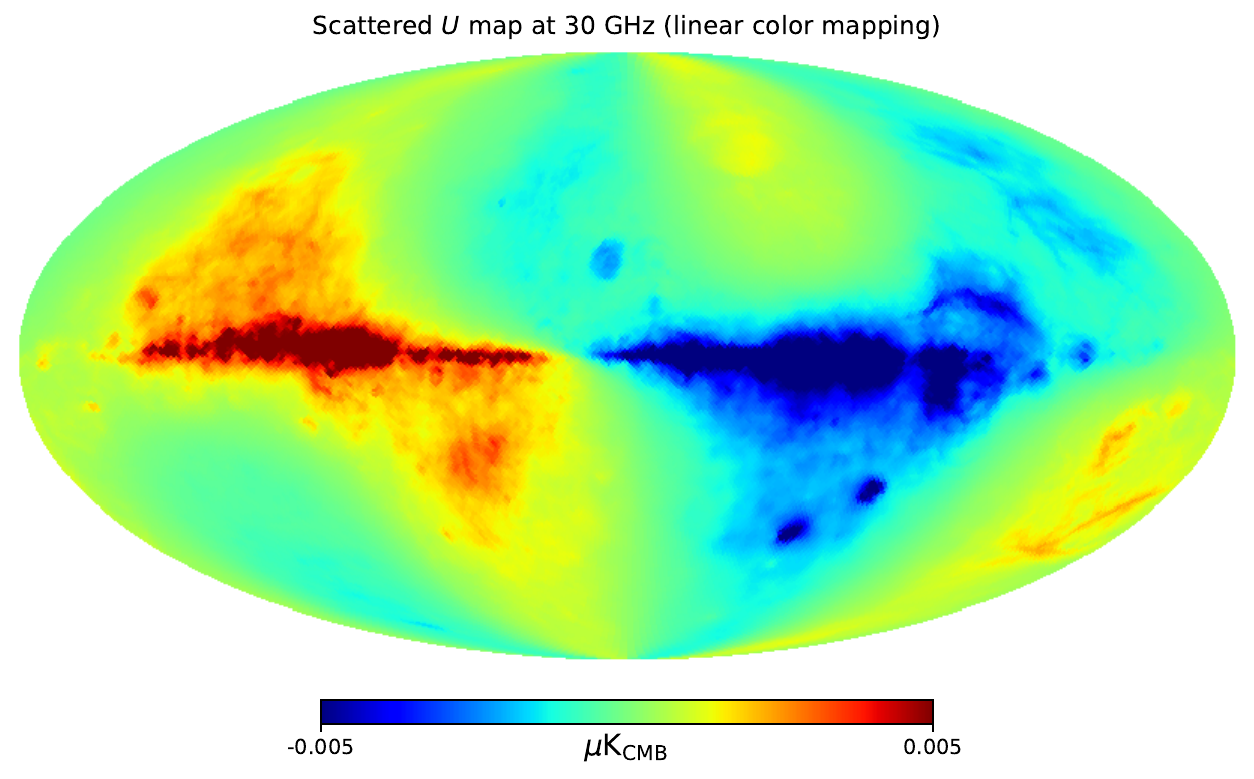}
\includegraphics[width=0.33\textwidth]{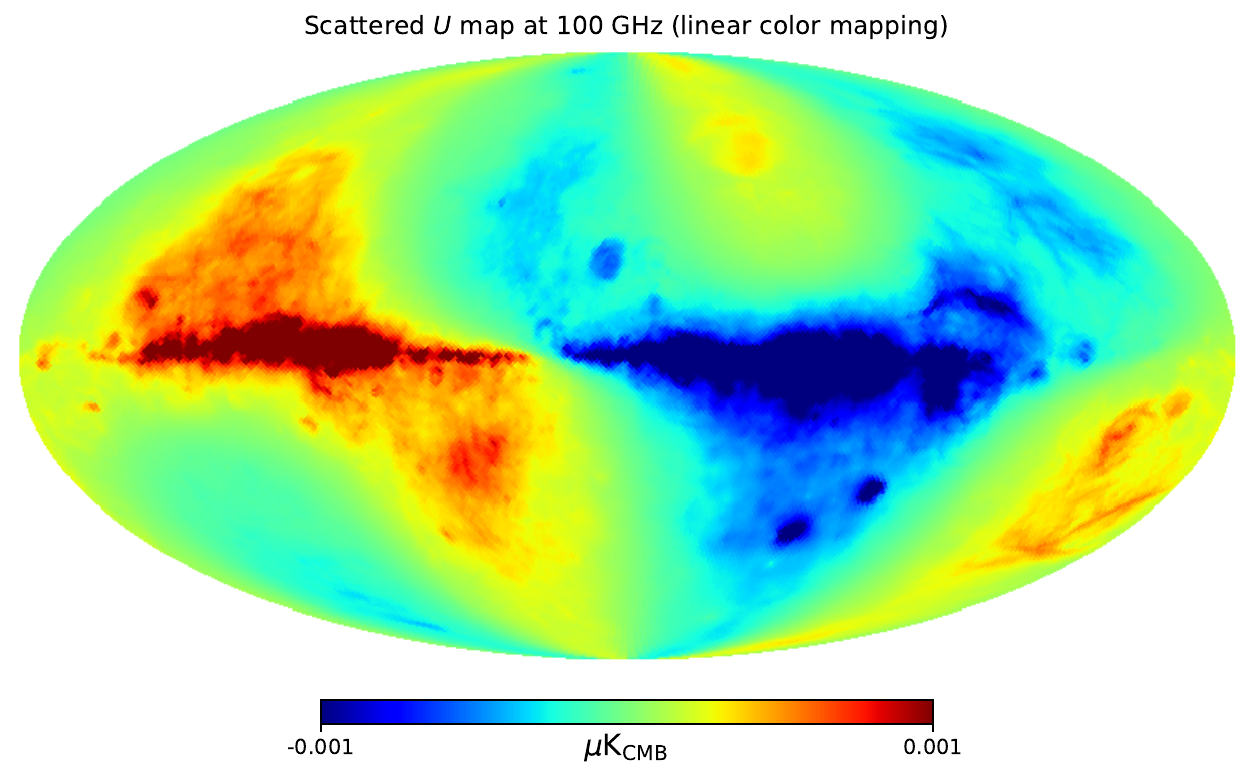}
\includegraphics[width=0.33\textwidth]{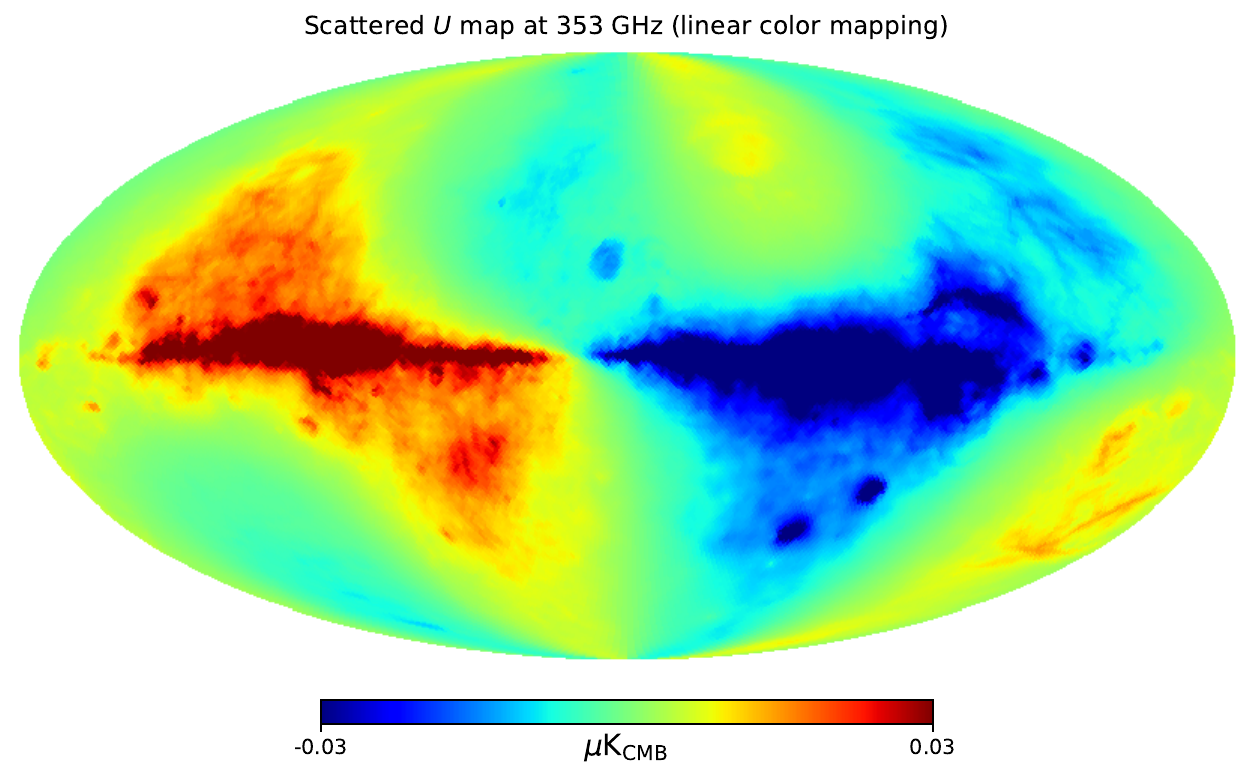}\\
\includegraphics[width=0.33\textwidth]{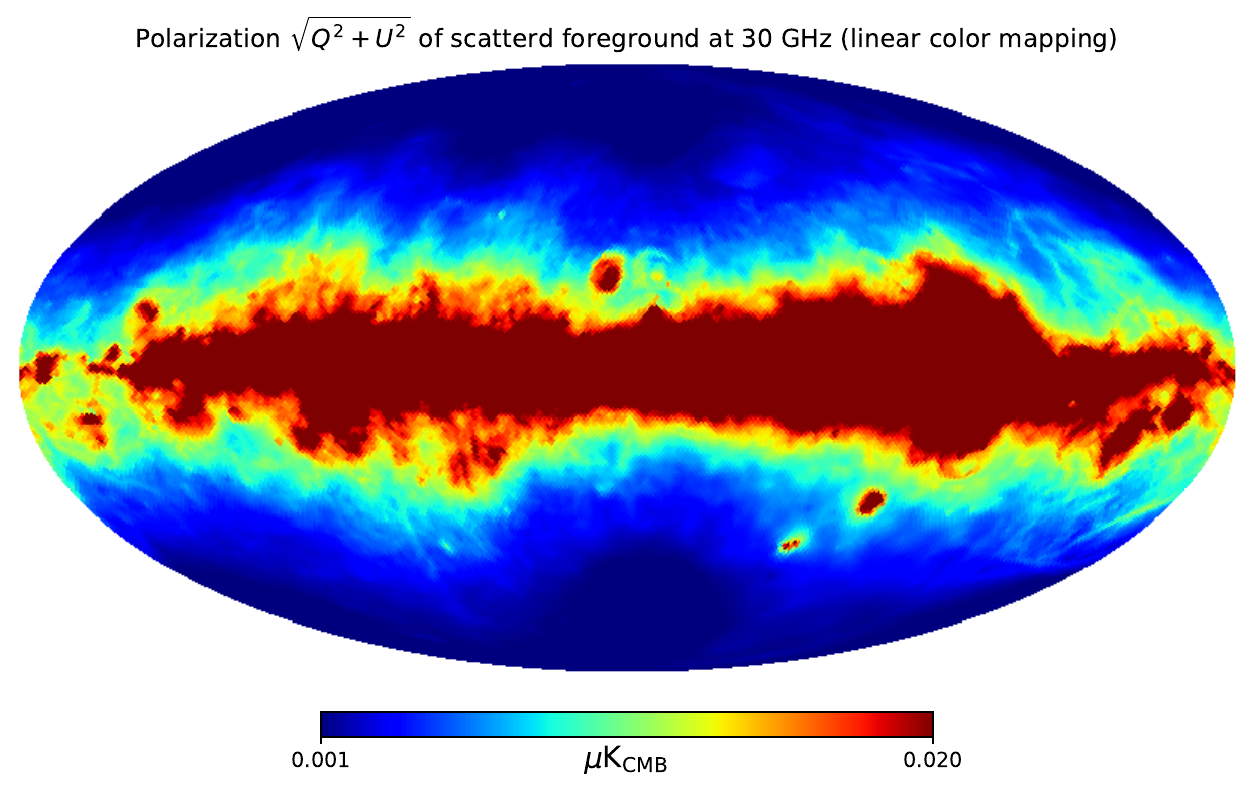}
\includegraphics[width=0.33\textwidth]{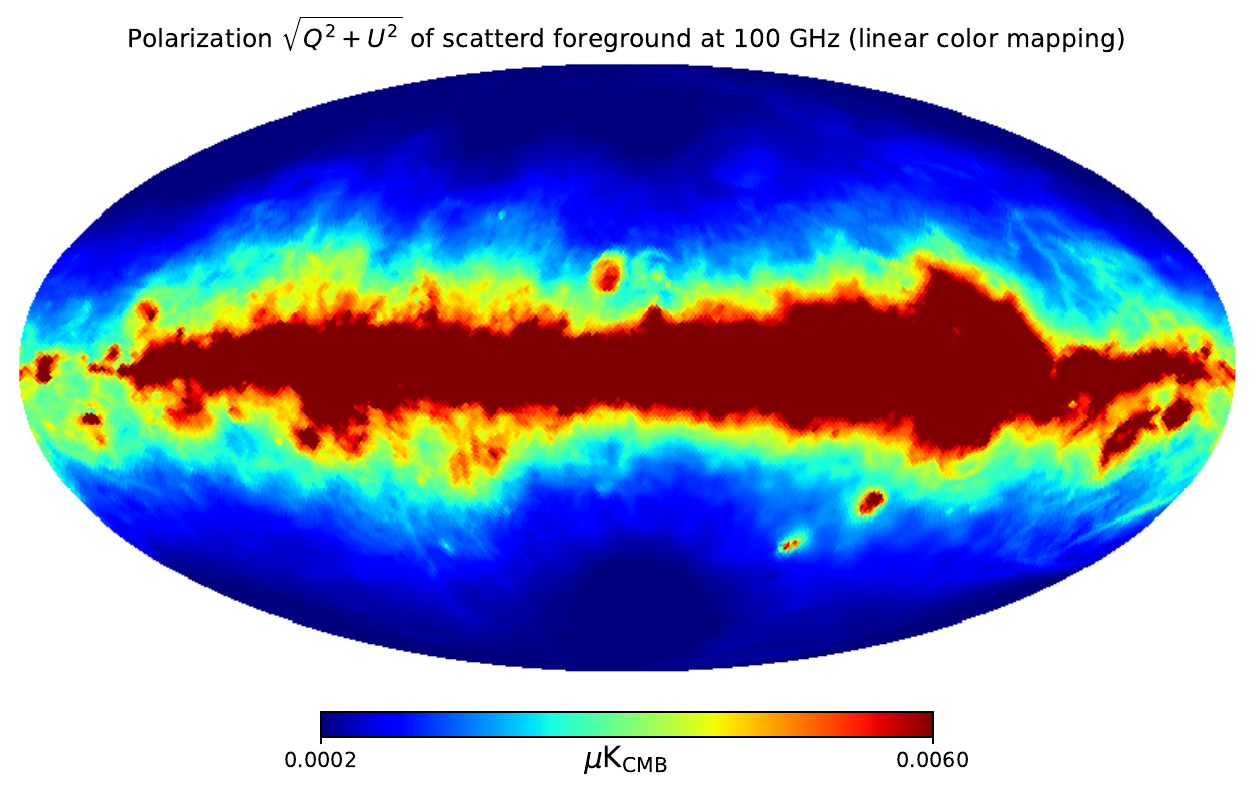}
\includegraphics[width=0.33\textwidth]{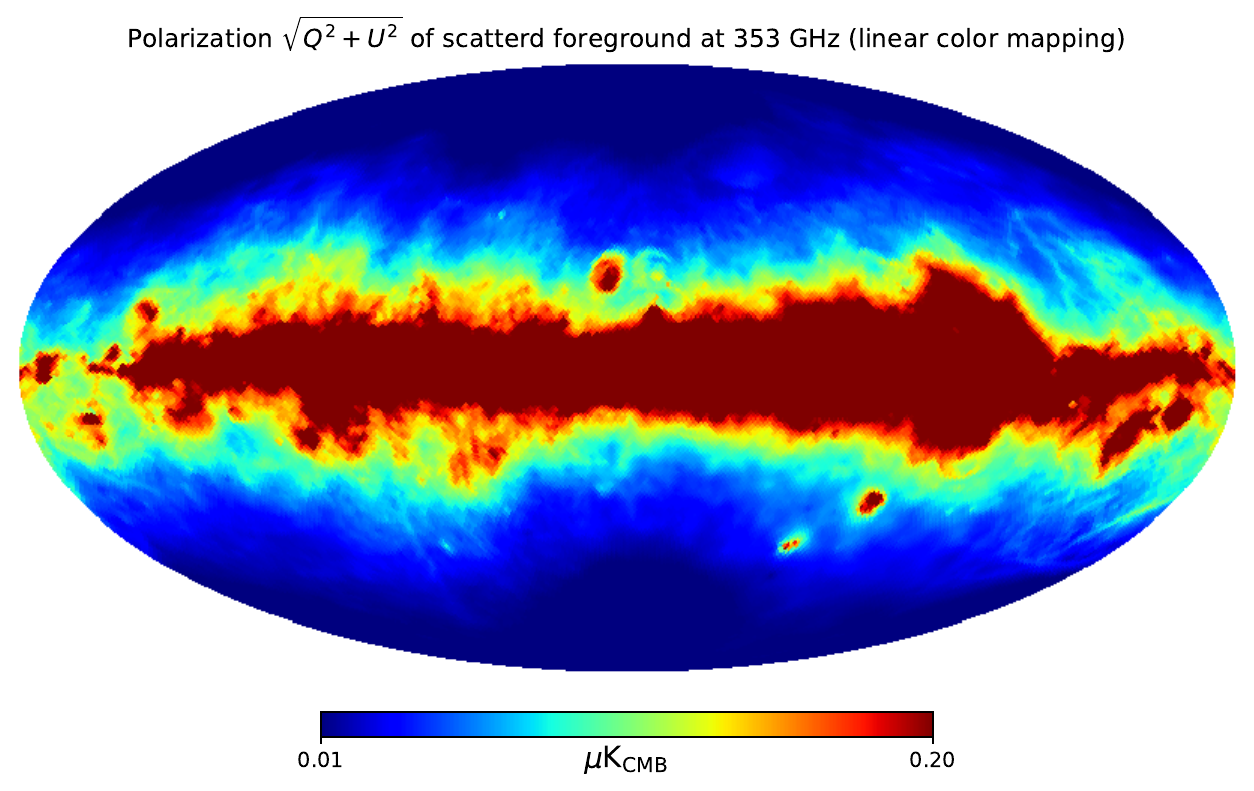}\\
\includegraphics[width=0.33\textwidth]{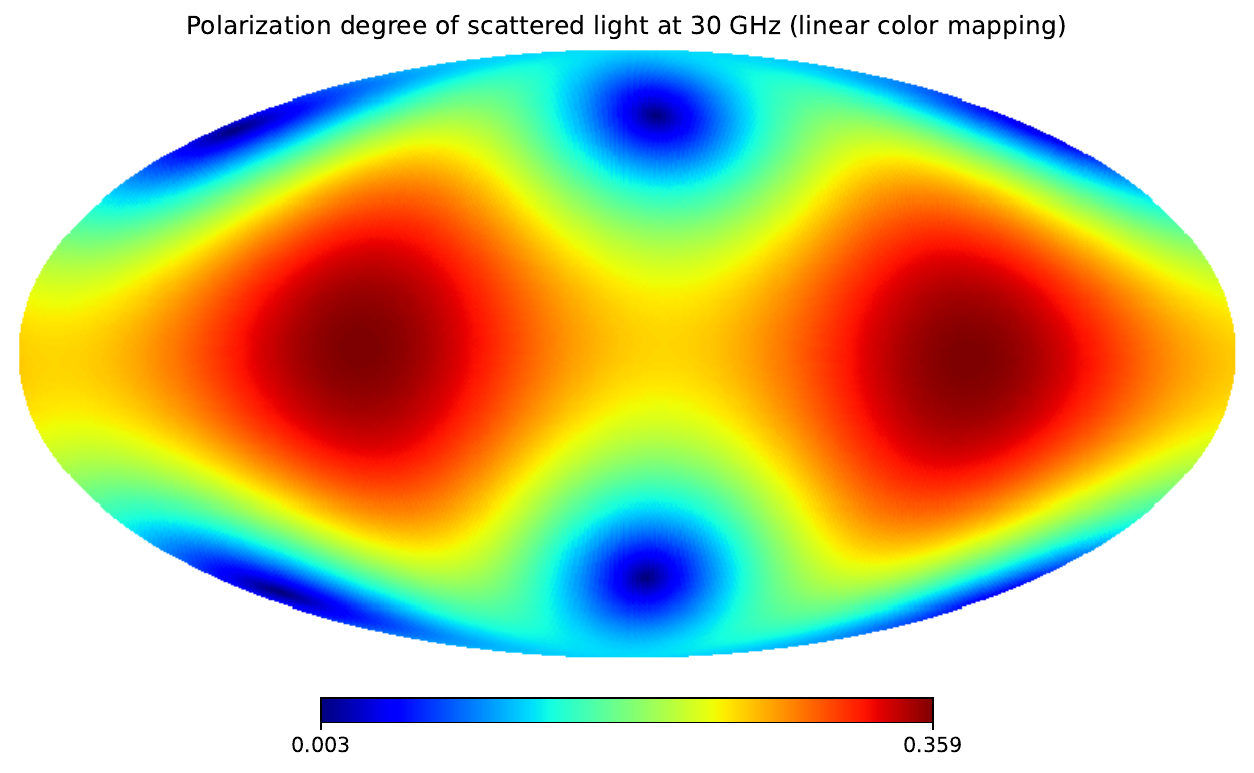}
\includegraphics[width=0.33\textwidth]{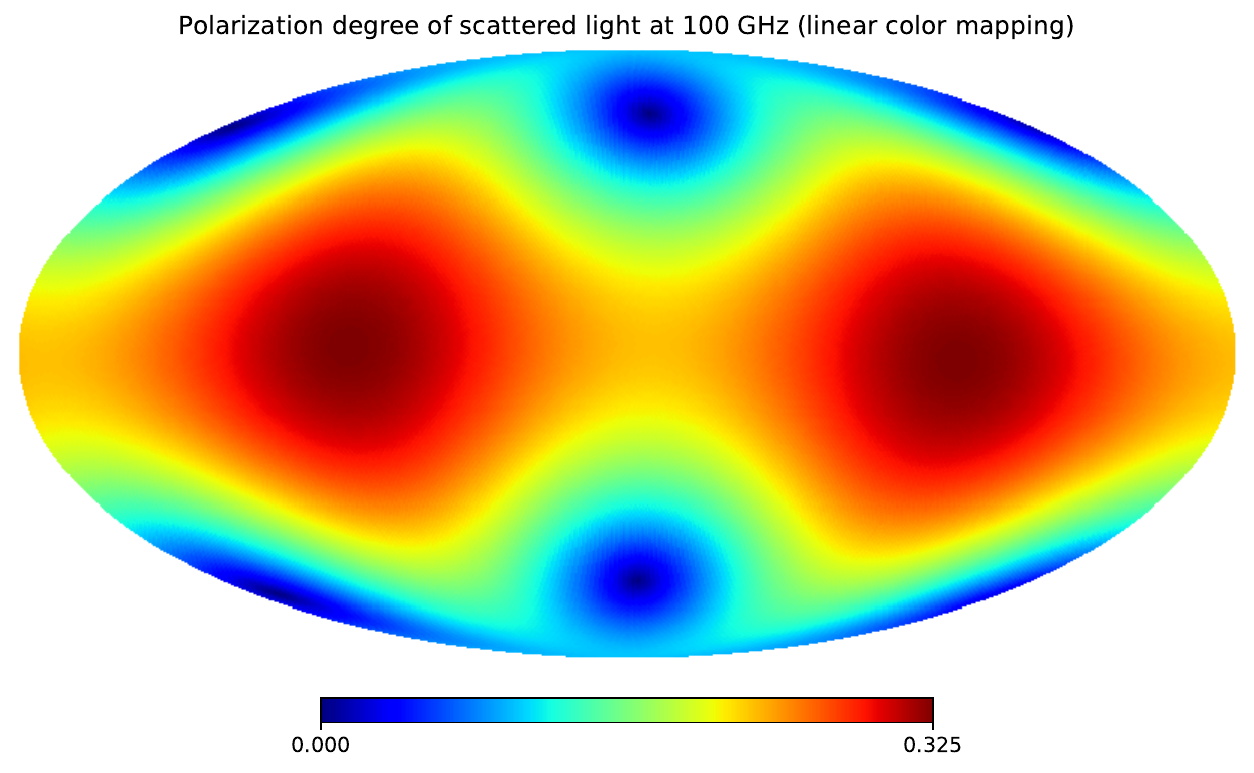}
\includegraphics[width=0.33\textwidth]{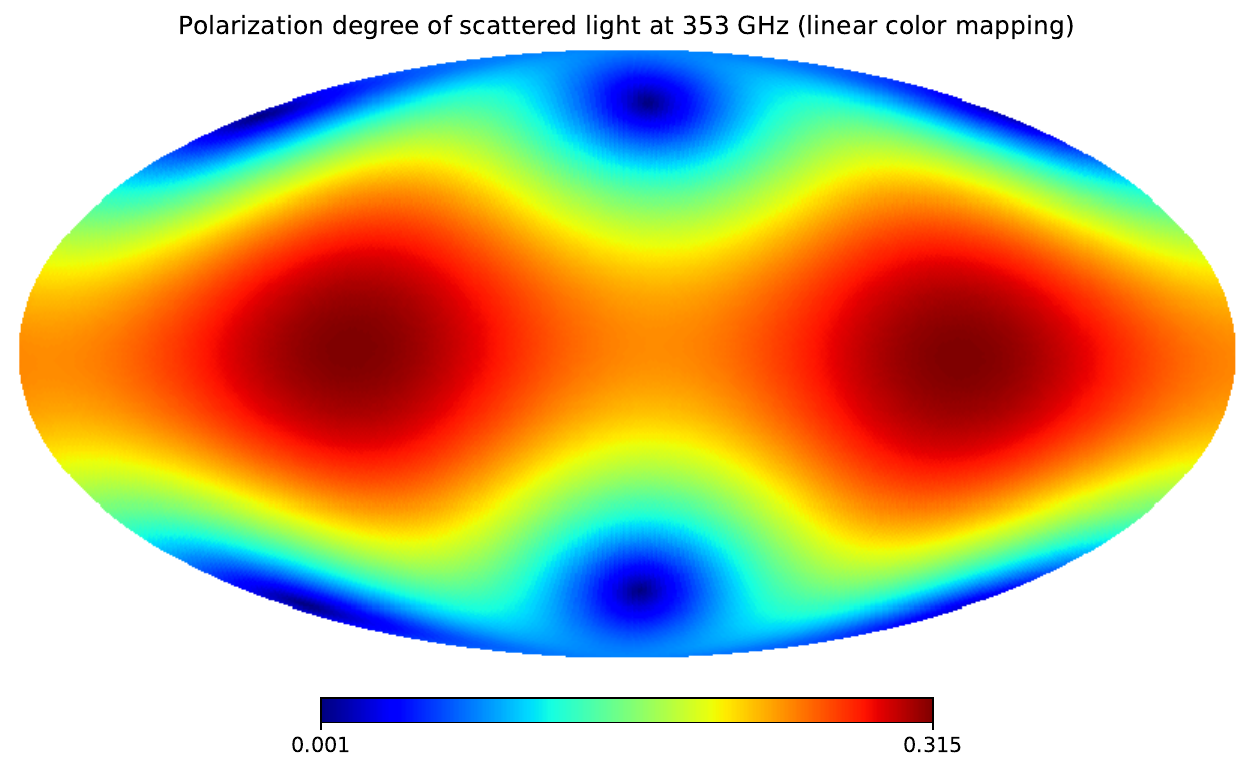}
\caption{From top to bottom: Stokes parameters $I$, $Q$, $U$, intensities of the polarization mode and polarization degrees of scattered foreground radiation by Galactic free electrons, in unit of $\mu\mathrm{K_{CMB}}$ (The dimension of polarization degree is unit-less). 
From left to right: maps at 30 GHz, 100 GHz and 353 GHz. 
For temperature mode, the map is in logarithmic colour mapping, and for others, in linear colour mapping. }
\label{Figure_Stokes_scattering}
\end{figure*}

\noindent Although \citet{2023arXiv230412350H} pointed out that the DMs in the inner disk may be under-estimated by a factor of about two, and also marginally underrated at high latitudes, we opted to use their DM map to estimate the order of magnitude of the signal originating from scattering by free electrons, with a specific interest in regions at mid to high Galactic latitudes.
We model incident radiation at the scatterers using \textit{Planck} observations with the CMB subtracted.\footnote{LFI\underline{ }CompMap\underline{ }Foregrounds-commander-030\underline{ }R3.00.fits,\\	
HFI\underline{ }CompMap\underline{ }Foregrounds-commander-100/353\underline{ }R3.00.fits}

In Fig. \ref{Figure_Stokes_scattering}, we display the resulting Stokes parameters and polarization degree $p = \sqrt{Q^2+U^2}/I$ of scatterd foreground emission by interstellar electrons, for three different frequencies of interest.
%, based on the YMW16 electron distribution model and electron density perturbations generated with a Kolmogorov spectrum. 

Unsurprisingly, the overall intensity of the emission closely follows the DM map that was used for this calculation. This is due to the approximation that each scattering electron ``sees" the same incoming radiation pattern. Away from the Galactic plane, the total amplitude of the scattered light is in the hundreds of nK range at 100 GHz, while polarized intensity is of a few nK. 

The polarization is dominated by the $Q$ Stokes parameter for obvious geometrical reasons: incoming radiation mostly arrives at the scatterers parallel to the Galactic plane, hence the scattered emission is polarized perpendicularly to the Galactic plane. The polarization degree reaches tens of percents at Galactic longitudes of 90 and 270 degrees.  

\begin{figure*}
\centering
\subfigure{\includegraphics[width=0.24\textwidth]{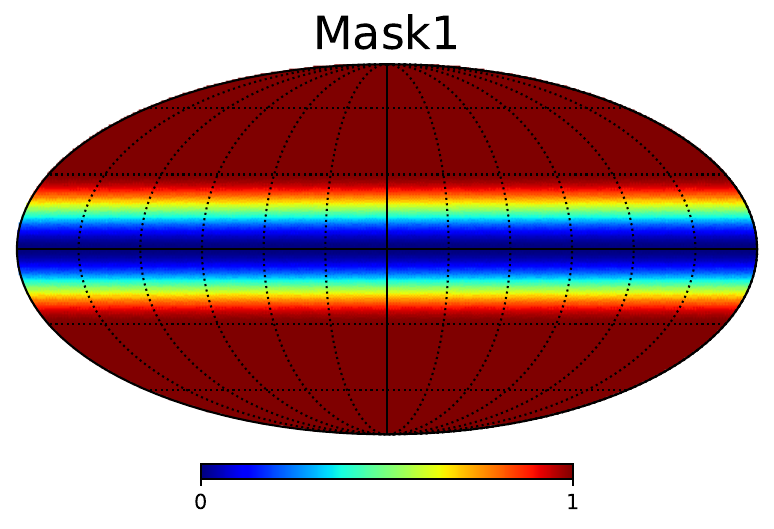}}
\subfigure{\includegraphics[width=0.24\textwidth]{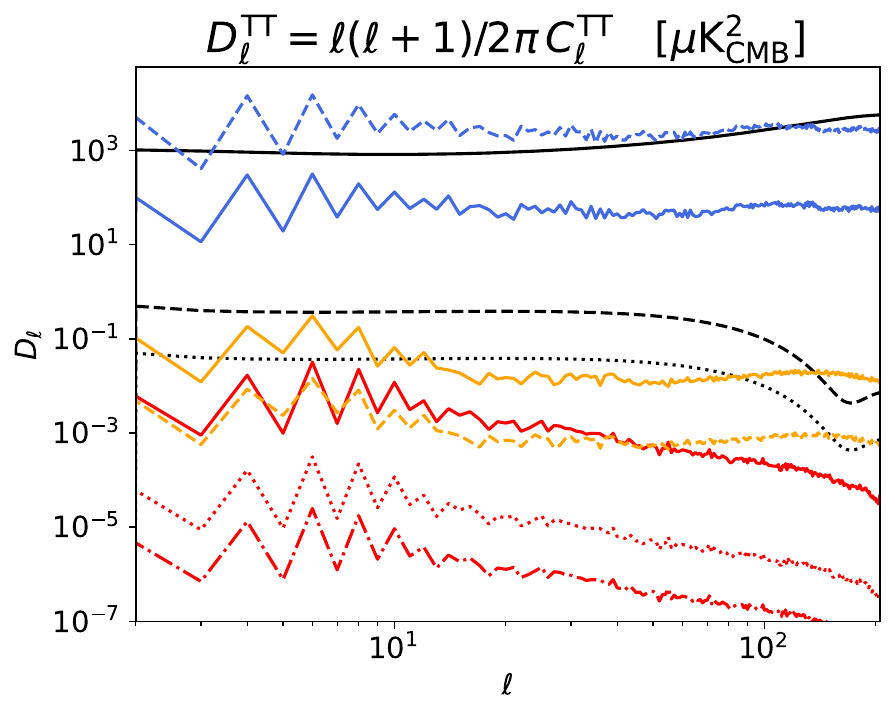}}
\subfigure{\includegraphics[width=0.24\textwidth]{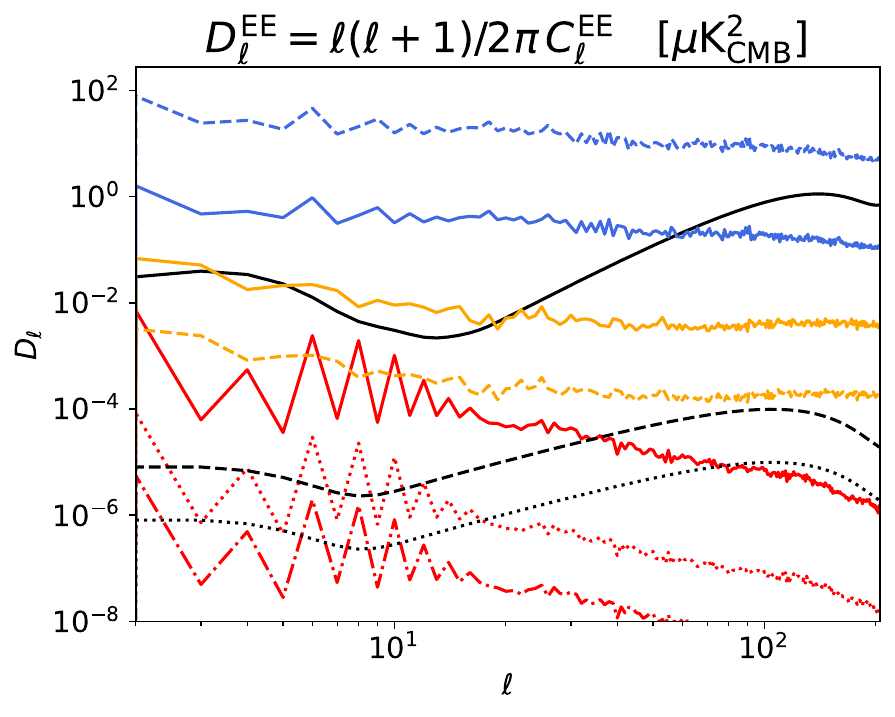}}
\subfigure{\includegraphics[width=0.24\textwidth]{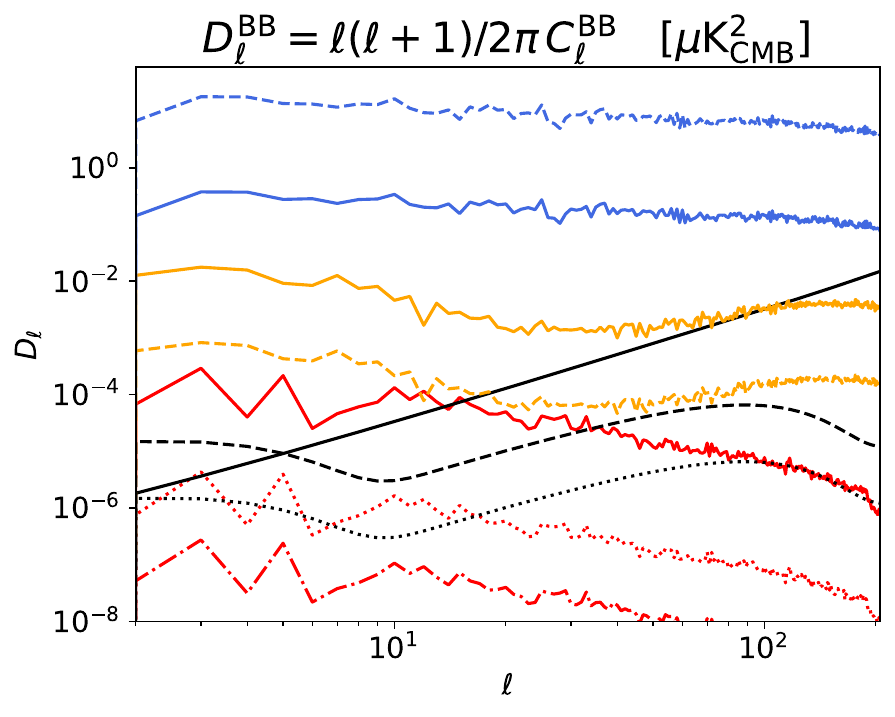}}

\subfigure{\includegraphics[width=0.24\textwidth]{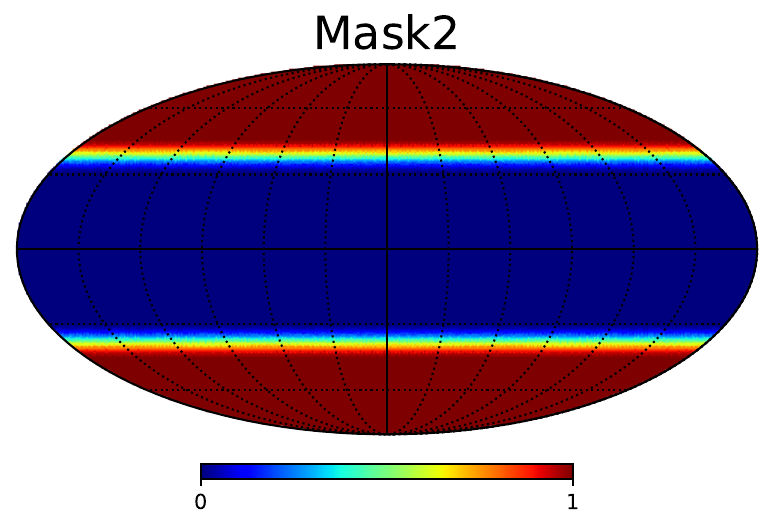}}
\subfigure{\includegraphics[width=0.24\textwidth]{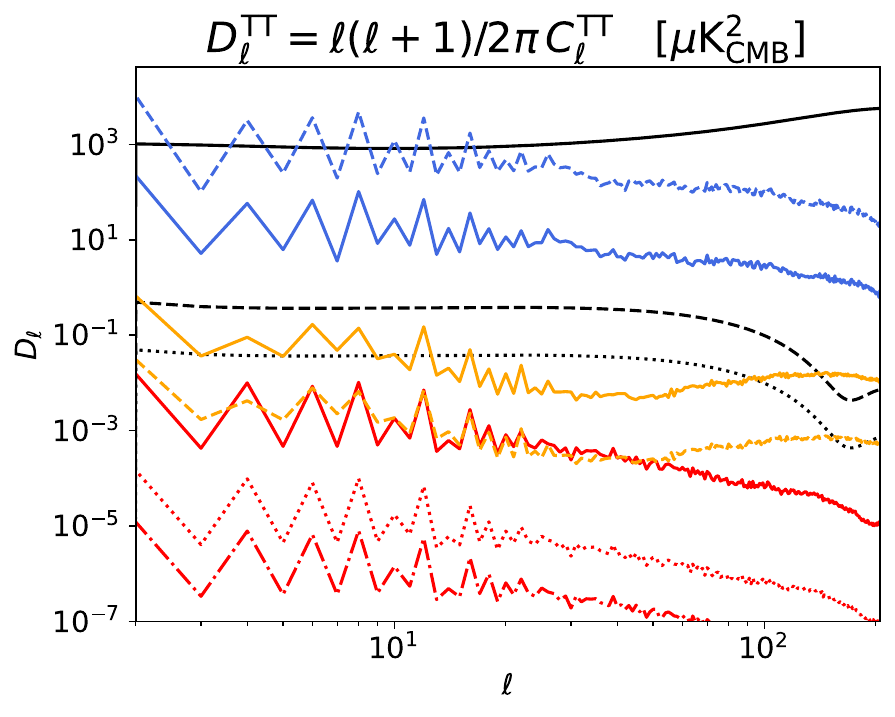}}
\subfigure{\includegraphics[width=0.24\textwidth]{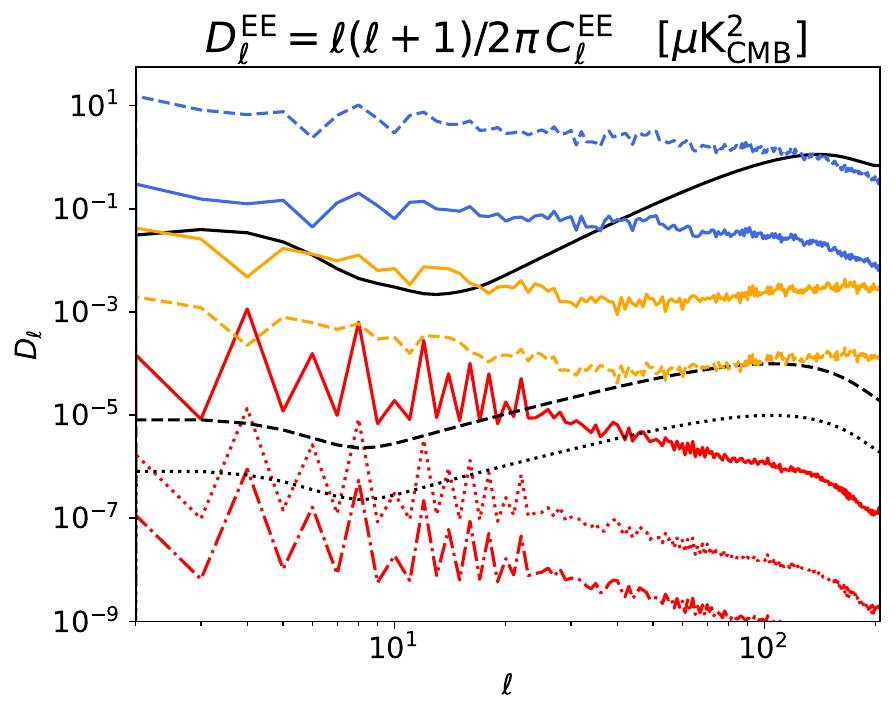}}
\subfigure{\includegraphics[width=0.24\textwidth]{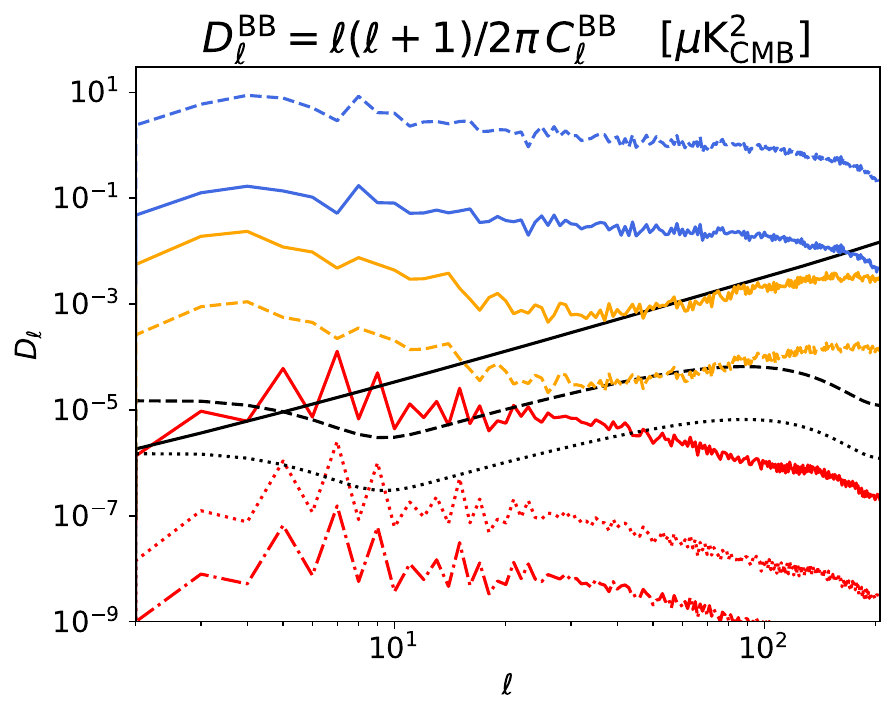}}

\subfigure{\includegraphics[width=0.98\textwidth]{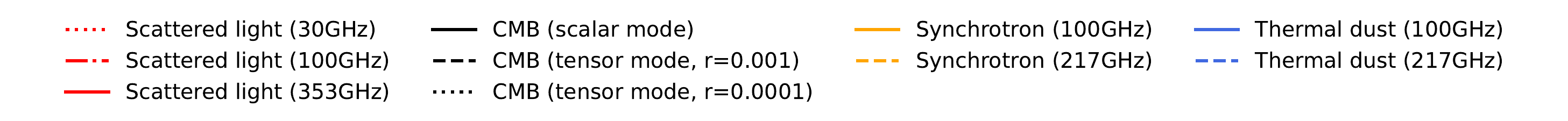}}
\caption{Angular power spectra $D_\ell^{TT}$, $D_\ell^{BB}$, $D_\ell^{EE}$ of scattered light at 30 GHz, 100 GHz, 353 GHz, primordial scalar mode CMB and primordial tensor mode CMB with tensor-scalar-ratio $r=0.001$ and $r=0.0001$, synchrotron emission and thermal dust emission at 100 GHz, 217 GHz.
Top: spectra calculated with Mask1. 
Bottom: spectra corresponding to Mask2. }
\label{Figure_DL_scattering}
\end{figure*}

We are particularly interested in the angular power spectra of scattered light in regions with low foreground emission, at medium and high Galactic latitude. We compute power spectra using two masks, displayed in Fig. \ref{Figure_DL_scattering}:
\begin{equation}
\mathrm{Mask1} = 
\begin{dcases}
1, & 0^\circ < \theta < 60^\circ, \, 120^\circ < \theta < 180^\circ\\
0.5 + 0.5\cos6\theta; & 60^\circ < \theta < 120^\circ
\end{dcases}
\end{equation}

\begin{equation}
\mathrm{Mask2} = 
\begin{dcases}
1, & 0^\circ < \theta < 45^\circ, \, 135^\circ < \theta < 180^\circ\\
0.5 - 0.5\cos12\theta, & 45^\circ < \theta < 60^\circ, \, 120^\circ < \theta < 135^\circ\\
0. & 60^\circ < \theta < 120^\circ
\end{dcases}
\end{equation}

Fig. \ref{Figure_DL_scattering} shows the comparison of the scattered Galactic foreground emission by free electrons, with scalar mode CMB, primordial tensor mode CMB with $r = 0.0001$ and $0.001$, synchrotron\footnote{We use COM\underline{ }CompMap\underline{ }Synchrotron-commander\underline{ }0256\underline{ }R2.00.fits map from \citet{planck2014-a12} as the map of Stokes I. 
The map of Stokes Q and U are those from \citet{2024arXiv240318123D}, which combined WMAP and \textit{Planck} low frequencies data, to obtain a low-noise full-sky map of synchrotron polarized emission. } and thermal dust foreground emission.\footnote{COM\underline{ }CompMap\underline{ }IQU-thermaldust-gnilc-varres\underline{ }2048\underline{ }R3.00.fits \citep{planck2016-l04}}
Although scattering of foreground is a small secondary effect, it is not completely negligible relative to primordial CMB tensor mode signals for $r$ values on the order of 0.001 or less. 
We note that even at higher latitudes, as displayed in the bottom part of Fig. \ref{Figure_DL_scattering}, this effect is still worth considering to some extent. 

Free electrons in our own Galaxy also scatter the CMB photons themselves, generating polarized emission by scattering the primordial CMB quadrupole off Galactic electrons. This effect, however, generates a contribution to sky emission that turns out to be very small, and can be safely ignored.

\section{Rayleigh scattering by ISM dust grains}
\label{Section_Rayleigh_scattering_by_ISM_dust_grains}

\begin{figure}
\centering
\subfigure{\includegraphics[width=0.4\textwidth]{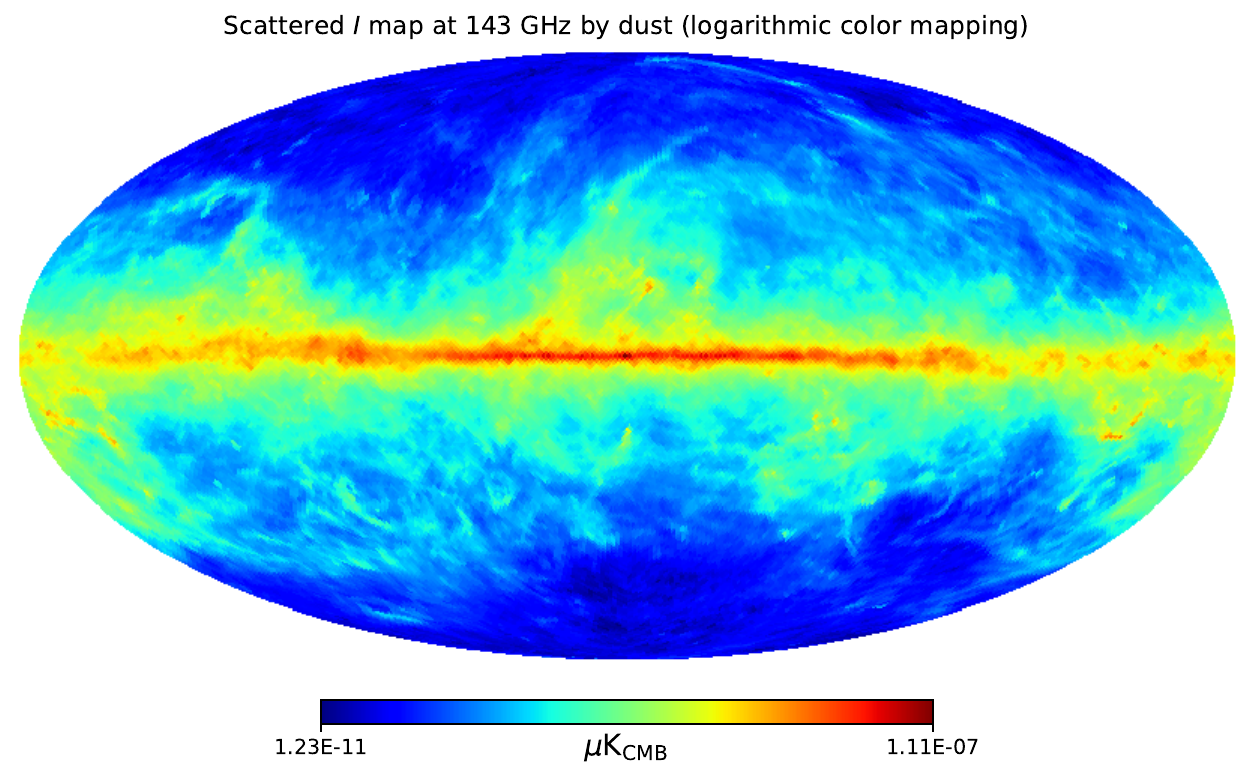}}

\subfigure{\includegraphics[width=0.4\textwidth]{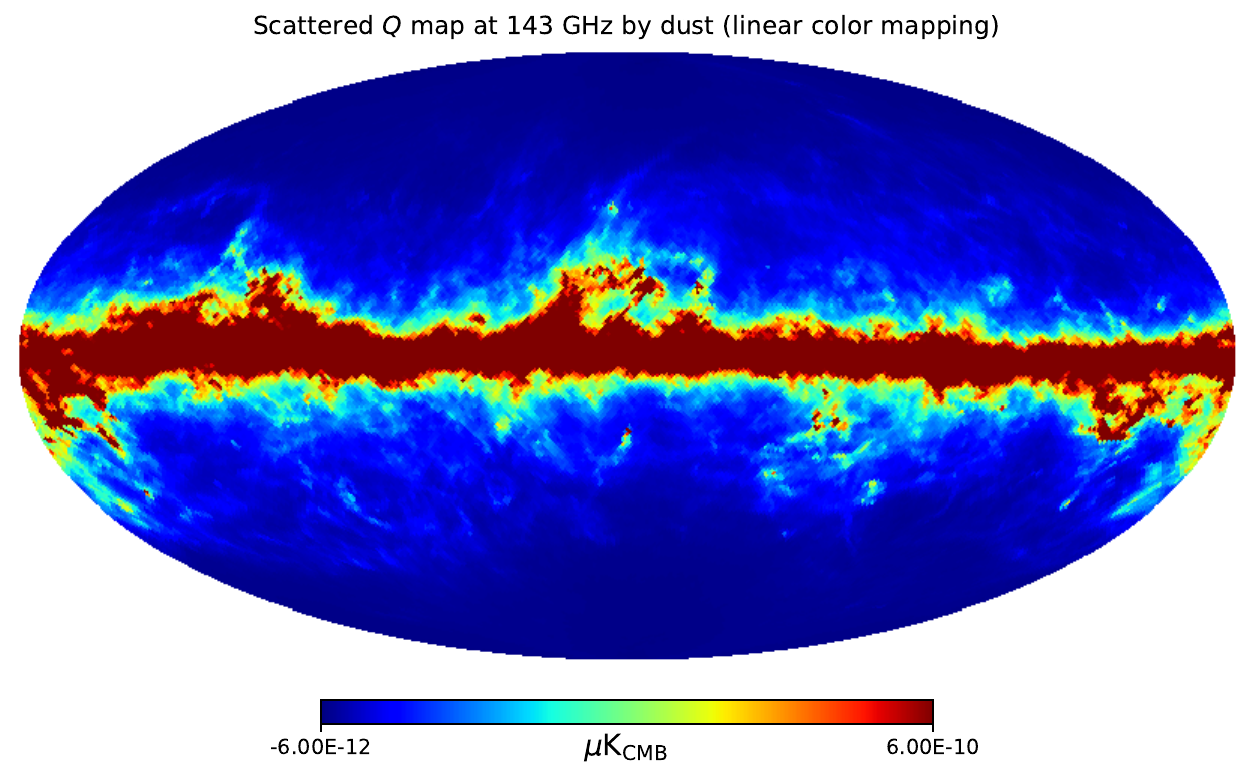}}

\subfigure{\includegraphics[width=0.4\textwidth]{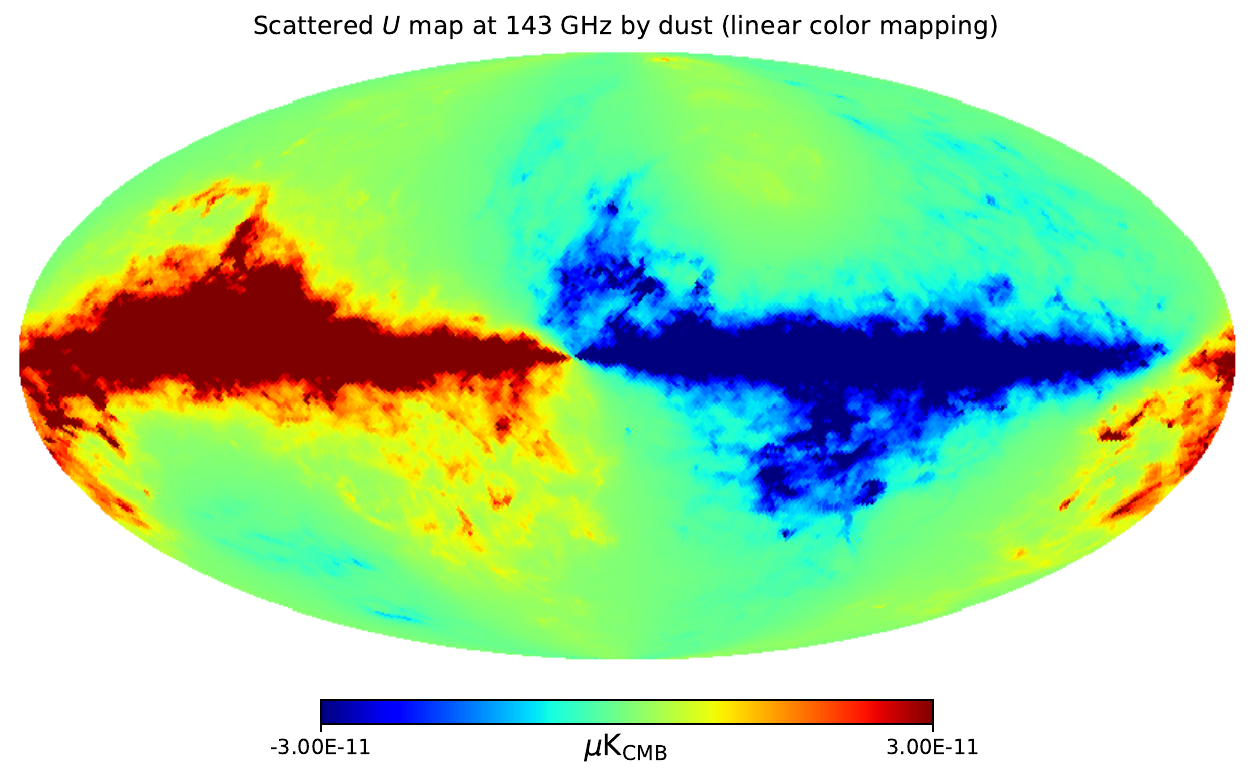}}
\caption{Maps for Stokes parameters of the scattered foreground by dust particles, at 143 GHz. 
From top to bottom: Stokes $I$, $Q$ and $U$}
\label{Figure_Stokes_scattering_by_dust}
\end{figure}

We now estimate the amplitude of scattering emission off dust grains in the ISM. 

The particle size distributions within ISM can be  approximated by power-law functions, as outlined by the model presented in \citep{1977ApJ...217..425M},
\begin{equation}
\mathrm{d}N_\mathrm{x} = \mathcal{A}_\mathrm{x}\, N_\mathrm{H}\, a^{-\beta}\mathrm{d}a, \quad a_{\min}<a<a_{\max}, \quad \beta\simeq3.3-3.6.
\end{equation}
For graphite particles, $0.005\,\mathrm{\mu m}<a<1\,\mathrm{\mu m}$; and for other materials, $0.025\,\mathrm{\mu m}<a<0.25\,\mathrm{\mu m}$. 
Specific constants $\mathcal{A}_\mathrm{x}$ are used for different materials: $\mathcal{A}_\mathrm{Si} = 10^{-25.10}\mathrm{cm}^{2.5}$ and $\mathcal{A}_\mathrm{C} = 10^{-25.13}\mathrm{cm}^{2.5}$ \citep{1984ApJ...285...89D}. 
Here, $N_\mathrm{H}$ represents the number density of hydrogen nuclei (encompassing both atomic and molecular forms) \citep{2001ApJ...548..296W} and accordingly, $\dd N_\mathrm{x}$ is the number density of dust grains with material ``x'', and size between $a$ and $a+\dd a$. 
However, it should be noted that, determining the precise number of large and small particles encounters challenges currently:
the former, characterised by a size $a > \lambda$ ($\lambda$ is for visible and infrared band), extinguish stellar light with negligible frequency dependence, a phenomenon known as "grey extinction". 
Since most methods for measuring dust extinction are related to reddening rather than direct extinction, they exhibit insensitivity to grey extinction \citep{2010MNRAS.401..231G}. 
In the latter case, the dust particles reside in the Rayleigh limit of the visible light band, resulting in an extinction cross-section that is proportional to the volume of the particles. 
Consequently, the extinction cross-section per unit mass remains size-independent, making it difficult to determine the size distribution of these smallest particles through extinction measurements \citep{1977ApJ...217..425M}. 

CMB observations mostly occur in the millimetre to centimetre wavelength range (30\;GHz to 300\;GHz), for which the size of dust grains $a$ smaller than $1/10\lambda$. This leads to Rayleigh scattering dominating the scattering interaction of CMB foreground emission with ISM dust. 
Though dust particles within ISM usually exhibit aspherical shapes \citep{2007JQSRT.106..225L}, it is reasonable to perform an order of magnitude estimate by modelling them as uniform, isotropic dielectric spheres with radius $a$ and relative permittivity $\epsilon_r = \Re{\epsilon_r} + \mathrm{i}\Im{\epsilon_r}$, 
where the imaginary component of the permittivity contributes to the dissipation of electromagnetic waves in dielectric materials. 
For naturally occurring dissipative media, $\Im{\epsilon_r} < 0$.\footnote{
    The sign of the imaginary part of permittivity depends on the Fourier transform convention. 
    In this paper we adopt the following one: 
    \begin{equation}
    f_\omega = \frac{1}{2\uppi } \int_{-\infty}^{+\infty} f(t) \mathrm{e}^{-\mathrm{i}\omega t} \mathrm{d}t. 
    \end{equation}
    }
The absorption cross-section for a single uniform isotropic dielectric sphere with volume $V$ is expressed as \citep{2019JAP...125j3105N}: 
\begin{equation}
\sigma_{\rm{abs}} = -\frac{24\uppi^2 a^3 \Im{\epsilon_r(\nu)}}{\Big(\Re{\epsilon_r(\nu)}+2\Big)^2 + \Big(\Im{\epsilon_r(\nu)}\Big)^2} \, \frac{\nu}{c},
\end{equation}
And the total scattering cross-section across all directions is described by \citep[Eq. (10.11)]{jackson_classical_1999}: 
\begin{equation}
\sigma_{\rm{sca}} = \frac{8\uppi}{3}\left(\frac{2\uppi\nu}{c}\right)^4 a^6 \left|\frac{\epsilon_r(\omega)-1}{\epsilon_r(\omega)+2}\right|^2.
\end{equation}

Similarly to the Thomson scattering case discussed in the previous section, with the distinction that the Rayleigh scattering cross-section depends on the incident wave's frequency, 
the polarization tensor of scattered light off single dust particle can be written as: 
\begin{equation}
J'_{0, ab} = \frac{3}{8\uppi} \frac{\sigma_\mathrm{sca}}{{r'}^2}\cdot\frac{2}{\mu_0 c} \int \Big\langle\big(\bm{E}\cdot\bm{e}'_a\big) \big(\bm{E}^*\cdot\bm{e}'_b\big)\Big\rangle \,\dd\Omega.
\end{equation}
Thus the total scattered light by all dust particles along one sight-line is: 
\begin{equation}
\begin{aligned}
J'_{ij}(\bm{n}') &= \frac{1}{\dd\Omega'}\int\dd\Omega'\int {r'}^2\dd r' \sum_{\mathrm{x=Si, C}}\int\dd N_\mathrm{x} J'_{0,ij}\\
&\simeq \frac{1}{3.5}\left(\frac{2\uppi\nu}{c}\right)^4\int N_\mathrm{H}\dd r'\sum_{\mathrm{x=Si, C}}\mathcal{A}_\mathrm{x}\left|\frac{\epsilon_r(\omega)-1}{\epsilon_r(\omega)+2}\right|^2 a_\mathrm{x,\max}^{3.5}\\
&\qquad\qquad\qquad\times\frac{2}{\mu_0 c} \int \Big\langle\big(\bm{E}\cdot\bm{e}'_a\big) \big(\bm{E}^*\cdot\bm{e}'_b\big)\Big\rangle \,\dd\Omega.
\end{aligned}
\label{Equation_1_Rayleigh_scattering}
\end{equation}
The dust responsible for interstellar extinction appears to be relatively well-mixed with the gas in the Milky Way \citep{draine2010physics}.
A well-established value exists for the mean ratio of total neutral hydrogen $\Sigma_{\rm{H}}$ to colour excess \citep{1978ApJ...224..132B}, given by: 
\begin{equation}
\frac{\Sigma_{\rm{H}}}{E(B-V)} = 5.8\times10^{21}\,\mathrm{atoms\,\,cm^{-2}\,mag^{-1}}.
\end{equation}
Lines of sight traversing diffuse gas in the Milky Way tend to have an average value of $R_V \equiv \dfrac{A_V}{E(B-V)}\simeq 3.1$ \citep{draine2010physics}. For $R_V = 3.1$, a connection can be established between the column density of hydrogen atoms and dust extinction at $V$ band, $A_V$: 
\begin{equation}
\frac{\Sigma_{\rm{H}}}{A_V} = 1.9\times10^{21}\,\,\mathrm{atoms\,cm^{-2}\,mag^{-1}}.
\end{equation}
Consequently, Eq. (\ref{Equation_1_Rayleigh_scattering}) can be related to the dust extinction data at $V$ band, yielding:  
\begin{equation}
J'_{ij}(\bm{n}') = \mathrm{ratio}\cdot\frac{2}{\mu_0 c} \int \Big\langle\big(\bm{E}\cdot\bm{e}'_a\big) \big(\bm{E}^*\cdot\bm{e}'_b\big)\Big\rangle \,\dd\Omega,
\label{Equation_2_Rayleigh_scattering}
\end{equation}
where
\begin{equation}
\begin{aligned}
\mathrm{ratio} &\simeq \frac{1.9\times10^{21}\,\mathrm{cm^{-2}\,mag^{-1}}}{3.5}\\
&\qquad\qquad
\times\left(\frac{2\uppi\nu}{c}\right)^4 A_V\left(\sum_{x=\mathrm{C, Si}}\mathcal{A}_\mathrm{x}\left|\frac{\epsilon_{\mathrm{x},r}-1}{\epsilon_{\mathrm{x},r}+2}\right|^2 a_{\mathrm{x},\max}^{3.5}\right).
\end{aligned}
\end{equation}
With $\nu = 100\,\mathrm{GHz}$ and $A_V \simeq 10\,\mathrm{mag}$, 
the ratio is approximately $10^{-12}$. This implies that only about $10^{-12}$ of the incident foreground emission will be scattered by ISM dust. 
Fig. \ref{Figure_Stokes_scattering_by_dust} shows the amplitude of the scattered foreground by interstellar dust at 143 GHz. 
It is reasonable to conclude that the scattering effect from ISM dust can be disregarded in the analysis of the CMB foreground emission.

\section{Conclusion}
\label{Section_Conclusion}

This paper addresses the computation of second-order effects in Galactic ISM emission arising from electric dipole scattering within the interstellar medium on the CMB foreground signals, which encompass both Rayleigh scattering by dust grains and Thomson scattering by thermal free electrons within ionized regions. 

The impact of scattering by interstellar dust particles on foreground signals is a relative correction of the order of $\sim 10^{-12}$. 
Consequently, the contribution of Rayleigh scattering can be safely disregarded in the analysis of CMB foreground contamination. 

Thomson scattering by free electrons, while still faint, is not obviously completely negligible for the most sensitive upcoming CMB experiments. This is due to the higher number density of interstellar free electrons as scatterers, the extreme sensitivity required for the detection of CMB polarization signals corresponding to the primordial tensor perturbations, 
and the high degree of polarization of scattered radiation.
It might be worth to refine our estimate in the future and possibly take the scattering of foreground emission by free electrons in the Milky Way into account in the next generation of CMB telescope programs aimed at searching for the trace of primordial gravitational waves. 

The spectral energy distribution of the scattered Galactic foreground emission by free electrons is identical to that of the incident emission, and its spatial distribution is highly correlated with free-free emission. 
Another key point is that the foreground emission scattered by free electrons is polarized, but free-free emission itself is not. One may try to detect the scattered emission by searching Q-type polarization with a mean ISM emission spectrum in mid to high Galactic latitude regions of ionized ISM.

We point out that the uncertainty of the work involved in this paper mainly comes from the spatial distribution of free electrons within the Milky Way, which is not very well measured as of now. 
However, 
based on the DM map from \citet{2023arXiv230412350H}, 
we could get a first-order estimate of the magnitude of the scattering effect, which may be refined in the future.

Our assumption that electrons as scatterers are all much closer to Earth than sources of incident foreground signals is another source of uncertainty.
Theoretically, a full three-dimensional model of the distribution of matter within the Milky Way would be necessary to refine our calculations.

\section*{Acknowledgments}
We thank Brandon Hensley, Chang Feng, V\'aclav Vavry\v{c}uk, Ya-Qi Zhao, and Rui Shi for helpful discussions and suggestions. 
This work is supported in part by National Key R\&D Program of China (2021YFC2203100), by NSFC (12261131497), by CAS young interdisciplinary innovation team (JCTD-2022-20), by 111 Project (B23042), by Fundamental Research Funds for Central Universities, by CSC Innovation Talent Funds, by USTC Fellowship for International Cooperation, by USTC Research Funds of the Double First-Class Initiative. Kavli IPMU is supported by World Premier International Research Center Initiative (WPI), MEXT, Japan.
All numerical calculations were operated on the computer clusters LINDA \& JUDY in the particle cosmology group at USTC.

\bibliographystyle{aa}
\bibliography{main}

\end{document}